\documentclass[
 reprint,showpacs,
superscriptaddress,
 amsmath,amssymb,
 aps]{revtex4-1}

\usepackage{graphicx}
\usepackage{dcolumn}
\usepackage{bm}
\usepackage{amsmath}
\usepackage{amsfonts}
\usepackage{amssymb}
\usepackage{sublabel}
\def\D{\mathrm{d}}

\begin{document}

\title{Spin-orbit controlled quantum capacitance of a polar heterostructure}

\author{Kevin Steffen}
\altaffiliation{Kevin.Steffen@physik.uni-augsburg.de} 
\affiliation{Center for Electronic Correlations and Magnetism, EP VI, Institute of Physics, University of Augsburg, 86135 Augsburg, Germany}
\affiliation{Max Planck Institute for Solid State Research, Heisenbergstra{\ss}e 1, 70569 Stuttgart, Germany}
\author{Florian Loder}
\affiliation{Center for Electronic Correlations and Magnetism, EP VI, Institute of Physics, University of Augsburg, 86135 Augsburg, Germany}
\affiliation{Center for Electronic Correlations and Magnetism, TP III, Institute of Physics, University of Augsburg, 86135 Augsburg, Germany}
\author{Thilo Kopp}
\affiliation{Center for Electronic Correlations and Magnetism, EP VI, Institute of Physics, University of Augsburg, 86135 Augsburg, Germany}

\begin{abstract}
Oxide heterostructures with polar films display special electronic properties, such as the electronic reconstruction at their internal interfaces with the formation of two-dimensional metallic states. Moreover, the electrical field from the polar layers is inversion-symmetry breaking and generates a Rashba spin-orbit coupling (RSOC) in the interfacial electronic system. We investigate the quantum capacitance of a heterostructure in which a sizeable RSOC at a metallic interface is controlled by the electric field of a surface electrode. Such a structure is, for example, given by a LaAlO$_3$ film on a SrTiO$_3$ substrate which is gated by a top electrode. Such heterostructures can exhibit a strong enhancement of their capacitance~\cite{Li}. The capacitance is related to the electronic compressibility of the heterostructure, but the two quantities are not equivalent. In fact, the transfer of charge to the interface controls the relation between capacitance and compressibility. We find that due to a strong RSOC, the quantum capacitance can be larger than the classical geometric value. However, in contrast to  the results of recent investigations~\cite{Grilli,Bucheli,Seibold} the compressibility does not become negative for realistic parameter values for LaAlO$_3$/SrTiO$_3$ and, therefore, we find that no phase-separated state is induced by the strong RSOC at these interfaces.
\end{abstract}


\date{\today}

\maketitle

\section{\label{sec:introduction}Introduction}
The static capacitance of a complex metallic structure is commonly considered to be a global quantity of the entire system and, consequently, is supposed to characterize the electronic state of the system rather insufficiently to deduce microscopic properties. However this is not always the case. For instance, recent experiments on polar oxide heterostructures with conducting LaAlO$_3$/SrTiO$_3$ interfaces suggest that the capacitance for these systems is a quantity which allows to draw distinct conclusions on the electronic interface state~\cite{Li,Tinkl}.  A special characteristic of this interface is that the electrons are subject to a sizeable spin-orbit coupling~\cite{Caviglia,Joshua,Zhong}. The question arises if spin-orbit coupling is a relevant control parameter for the capacitance of the interface. 

\begin{figure}[b]
\centering
\includegraphics[width=0.9\columnwidth]{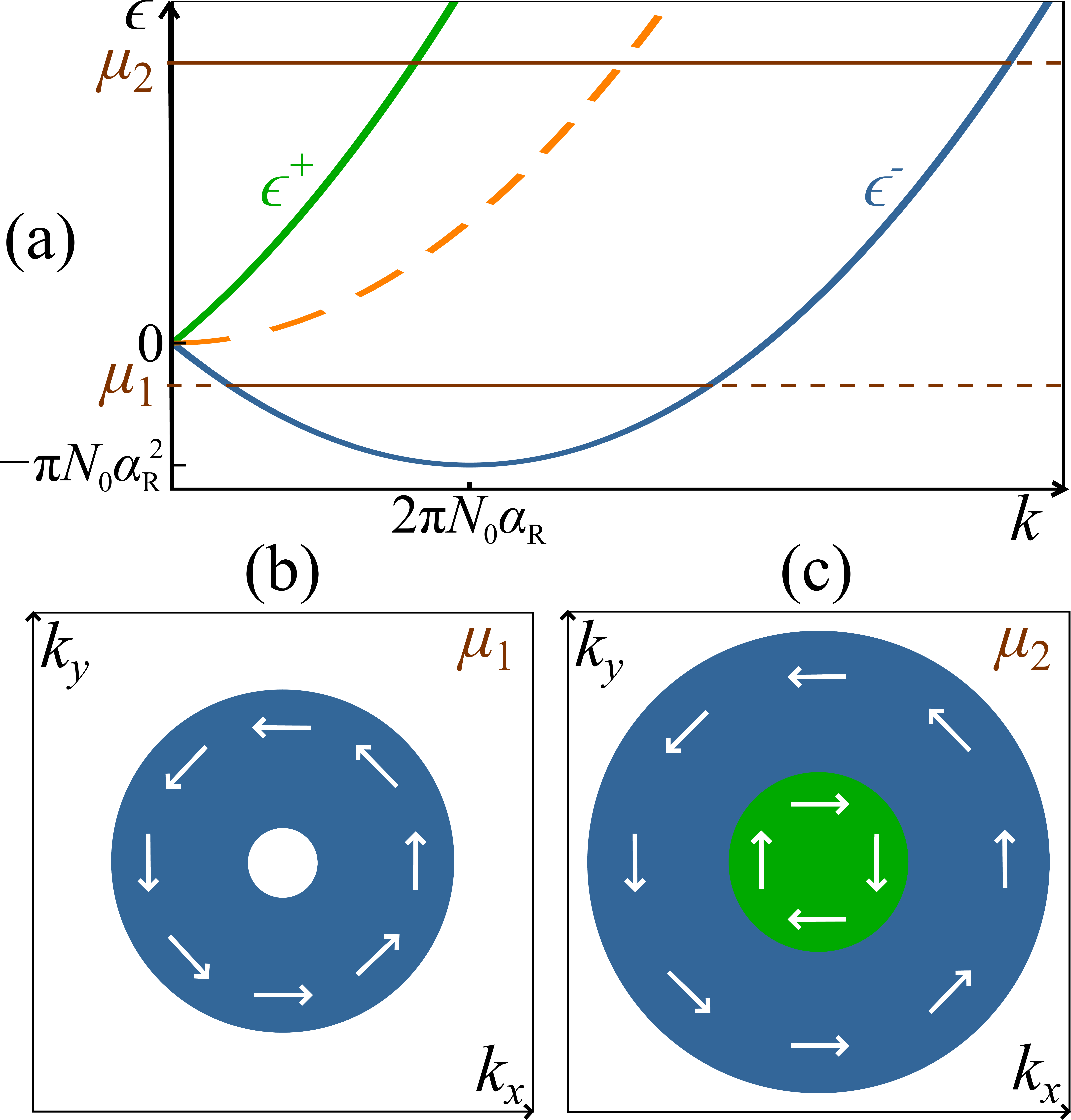}
\caption{(Color online) \textit{Top:} Dispersion of a planar electron system with RSOC. The dashed line indicates the dispersion for zero Rashba coupling. \textit{Bottom:} Occupation in $k$-space for $\mu_1<0$ (left, only the lowest band is occupied) and for $\mu_2>0$ (right, both bands are occupied).}
\label{Dispersion}
\end{figure}

Here, we address the fundamental issue whether spin-orbit coupling of the interface electrons allows to modify the capacitance of the heterostructure significantly, and we estimate the spin-orbit contribution for systems based on LaAlO$_3$/SrTiO$_3$ interfaces.
An enhancement of the capacitance is possible in the presence of a sufficiently large Rashba-like spin-orbit coupling (RSCO), because the RSOC allows to alter the electronic band energies by controlling the charge carrier density.
Such a RSOC appears to explain magnetotransport measurements at LaAlO$_3$/SrTiO$_3$ interfaces~\cite{Caviglia,Shalom,Liao,Hernandez,Joshua}, and it has been shown within {\it ab initio} evaluations that a RSOC is generated by the electric field of the polar LaAlO$_3$ layers on top of the SrTiO$_3$ substrate~\cite{Zhong}. The RSOC lowers the band edge quadratically with the spin orbit coupling parameter $\alpha_{\rm R}$ (see Fig.~\ref{Dispersion}). This even allows for a negative compressibility assuming that $\alpha_{\rm R}$ depends on the  electric field $E$ of the polar LaAlO$_3$ layers which in turn is related to the charge density at the LaAlO$_3$/SrTiO$_3$ interface~\cite{Grilli,Bucheli}. With their spectacular result of a negative compressibility in the presence of a sizeable RSOC, the authors of Refs.~\cite{Grilli,Bucheli,Seibold} suggest that this provides an intrinisic mechanism for the inhomogeneous phases observed at LaAlO$_3$/SrTiO$_3$ interfaces---yet, other scenarios for phase separation at LaAlO$_3$/SrTiO$_3$ interfaces have also been investigated recently~\cite{Ariando,Pavlenko}.

It is well known, that a negative electronic compressibility on one or both electrodes of a capacitor can lead to an enhancement of the capacitance beyond its geometrical value~\cite{Eisenstein,Kopp}. In fact, the geometric capacitance of a two-plate capacitor, $C_{\rm geom}= \varepsilon A/4\pi d$, is modified by electronic contributions which are taken into account in the electronic compressibility of the plates  $\kappa_i=(n_i^2\, \partial \mu_i/\partial n_i)^{-1}$, the derivative of the charge carrier density $n_i$ at plate $i$ with respect to the chemical potential $\mu_i$. Here $\varepsilon$ is the dielectric constant of the material between the two electrodes of area $A$, and $d$ is the thickness of the dielectric.
The kinetic contributions to the capacitance were coined by Luryi~\cite{Luryi} as quantum capacitance. This notation may be extended to all electronic compressibility terms (see Ref.~\onlinecite{Kopp}).

Already in the early nineties, a capacitance enhancement of semiconductor heterostructures has been identified  when the electronic compressibility of quantum Hall systems was investigated~\cite{Eisenstein}. The enhancement of the semiconductor capacitance could be traced quantitatively to the presence of quantum exchange effects in the electrodes. The contribution from electronic correlations to the capacitance is not negligible for very dilute electron systems~\cite{Kopp} and for electronic systems close to half-filling~\cite{Freericks}. However, the charge density at LaAlO$_3$/SrTiO$_3$ interfaces is in the intermediate range, neither as low as in the semiconductor systems~\cite{Eisenstein} nor close to half-filling. Therefore, the more sizeable capacitance enhancement of the oxide LaAlO$_3$/SrTiO$_3$ heterostructure~\cite{Li} and its negative compressibility~\cite{Tinkl} cannot be explained exclusively by exchange and correlation effects and other mechanisms might become relevant. 

To determine the thermodynamic electronic compressibility, two alternative procedures are conceivable: (i) one can either consider the interface subsystem with an {\it external} electric field which induces the RSOC. In order to identify the electronic compressibility of the interface electrons, one has to keep the electric field fixed in thermodynamic derivatives. Or (ii), one evaluates the compressibility for the entire electronic system, for which the RSOC-generating field is an {\it internal} field. 

A scheme in the sense of (i) has been suggested by the authors of Ref.~\cite{Grilli}.  There the electric field was set as a function of the interface charge carrier density $n$ and, then, the inverse compressibility was identified from the derivative of the chemical potential $\mu$ with respect to $n$, keeping any dependence else fixed. We disagree with this approach because there the electric field has to be kept fixed and therefore does not contribute to the compressibility~\cite{comment1}.  In this paper we establish the second procedure (ii) and relate the electronic compressibility of the entire electronic system directly to the capacitance of the device---consisting of the interface, dielectrics, and the surface of the LaAlO$_3$ film or a metallic top electrode. 

As indicated above,  previous investigations focused on the interface subsystem~\cite{Grilli,Bucheli}, whereas we consider the entire system. For realistic values of the RSOC, we identify a positive compressibility for the most part of the parameter space indicating that the system is thermodynamically stable. Nevertheless, we find an increased capacitance with respect to its geometrical value, the capacitance enhancement being caused by the RSOC in the metallic interface. 

In this article, we will investigate two different models, both building on a two-dimensional (2D) metallic system in which a RSOC modifies the electronic state: First we analyze a capacitor with two metallic planes, where the top electrode is a 2D or 3D metal (gate) and the second plane is the interface between a polar and non-polar insulator. The interface attains a metallic state with RSOC upon charge transfer from the top plate to the interface and the concomitant build-up of an internal electric field. Finally we study a two-layer system with the spin-orbit coupling controlled by the filling of three $t_{2g}$-orbitals at the interface. We will show that the multilayered system has a range of parameters where it has a positive total compressibility, but its capacitance is enhanced above its geometrical value on account of the spin-orbit coupling in the metallic interface plane. In this analysis we adjust the parameters to those specific for LaAlO$_3$/SrTiO$_3$ (LAO/STO) heterostructures in order to explore the feasibility of spin-orbit controlled capacitance enhancement with a polar dielectric material.

\section{\label{sec:basics}Rashba Spin Orbit Coupling and Compressibility}

In a 2D free electron system subject to a Rashba spin-orbit coupling (RSOC) $\gamma({\bf E}\times{\bf p})\cdot{\bf S}$, with coupling constant $\gamma$, the dispersion $\epsilon^\pm({\bf k})$ is given by (see, for example, Ref.~\cite{Winkler}):
\begin{align}\label{rsocdispersion}
\epsilon^\pm({\bf k})=\frac{k^2}{2\pi N_0}\pm\alpha_{\rm{R}} k,
\end{align}
where $N_0=m_0/\pi\hbar^2 =(m_0/m_e)/(\pi a_{\rm B}e^2)$ is the density of states for both spin directions in the absence of RSOC, $\alpha_{\rm R}$ defines the strength of the RSOC, and $m_0$ is the effective mass of the charge carriers. The RSOC is made possible by the absence of inversion symmetry, caused by an electric field perpendicular to the two-dimensional layer containing the electrons with areal density $n$. The spin-orbit splitting of the dispersion, Eq.~(\ref{rsocdispersion}), is shown in Fig.~\ref{Dispersion}. If $\mu<0$, only the lower band is occupied and the Fermi surface is an annulus; for $\mu>0$ the two occupied bands yield two concentric circles in $k$-space. As the lowest energy $\epsilon^{-}_{\rm min}=-\pi N_0\alpha_{\rm{R}}^2$ of the dispersion is proportional to the strength of the RSOC, it decreases with increasing strength of the symmetry breaking electrical field. Thus a stronger electrical field lowers the chemical potential for constant electron density $n$. In Ref.~\cite{Grilli} it was suggested that a negative compressibility can be obtained from exploiting this effect.

The inverse compressibility is obtained by the second derivative of the free energy $F$ per area with respect to the electron density \textit{with all external variables held fixed}. Since the symmetry breaking electrical field (SBEF) is an external variable in this scheme (and the field generated by the electrons in the plane themselves is symmetric), $\alpha_{\rm R}$ is independent of $n$ in the thermodynamic derivatives, and there is no dependence of the compressibility on the RSOC. 

Here we propose an extended scheme in which the electric field from the polar film, and consequently the RSOC, is an internal parameter, i.e., a quantity that depends on the electron density $n$. Then the density dependence of $\alpha_{\rm R} $ has to be accounted for when taking the derivative of the energy for the compressibility. In addition to the interface layer with a finite RSOC (L0), a second layer (L1), e.g., the surface of the LAO film in the LAO/STO heterostructure, comprising electrons and a positive background charge, is placed parallel at a distance $d$. Eventually, both layers---each of them electrically neutral---are connected. Since typically both layers had different electrochemical potentials, electrons will be exchanged between L0 and L1 until both layers are on the same electrochemical potential. It is of no relevance for the static capacitance how electronic charge is transferred between the plates. However, an electric field, which is proportional to the transferred amount of charge, is generated. This internal field breaks the inversion symmetry and thus generates RSOC in the respective layer. 

It stands to reason that also the electrons in the surface layer L1 experience a RSOC because the same electric field acts on the electronic states in L1. In the case of a bare RSOC  $\gamma=\mu_{\rm B}/\hbar m_e c$ in  $\gamma\,({\bf E}\times{\bf p})\cdot{\bf S}$ for free electrons with mass $m_e$, momentum ${\bf p}$ and spin ${\bf S}$ in a plane subject to a transverse electrical field ${\bf E}$, a reasoning demanding an equivalent treatment of layers L0 and L1 with respect to the RSOC appears to be appropriate. However, the bare RSOC with electric fields generated by the charged layers is orders of magnitude too small to have sizeable impact on the electronic reconstruction in the heterostructures. The resolution of this apparent unfeasibility of bare RSOC effects has already been established for semiconductor heterostructures (see the book by R.~Winkler~\cite{Winkler} for a comprehensive treatment) and is implemented for the LAO/STO interfaces as follows~\cite{Joshua,Zhong}: The Ti-ions provide an atomic SOC, whereas an internal electric field which displaces the Ti-ions slightly out of their in-plane positions, couples the three $t_{2g}$ orbitals. As spin and orbital space couple, one has to diagonalize six bands. The diagonalization of the 6-band system generates a SOC of Rashba type $\pm \alpha_{\rm R} k$ (see Eq.~(\ref{rsocdispersion})) even though only the lowest of the $t_{2g}$ bands (the $d_{xy}$) may be partially occupied. The RSOC parameter $\alpha_{\rm R}$ has to be estimated from electronic structure evaluations~\cite{Zhong}. In fact, from magnetotransport measurements, a sizeable RSOC has been identified for the LAO/STO interface electronic state ($\alpha_{\rm R}^{\rm ex}\approx (1\dots 5)\times 10^{-10}e{\rm Vcm}\approx (0.74\dots 3.7)\times 10^{-3}e^2$ for different backgate voltages), where the value of $\alpha_{\rm R}^{\rm ex}$ depends on the charge carrier density which is controlled by the internal electric field. In Sec.~\ref{sec:3band} we will address the 6-band situation explicitly. Until then we consider an effective two-band situation with a RSOC that is controlled by the internal field that depends in turn on the charge density. The actual set-up for the electronic surface states (layer L1) is very different and SOC effects have so far not been found to be relevant. Therefore we disregard RSOC in the surface layer L1 in our  approach.
 
Through the mechanism outlined above, the strength of the induced RSOC, $\alpha_{\rm R}$, becomes density dependent: We denote the Helmholtz free energies of the single layers as $F_0(n)$ and $F_1(n_1)$, calculated in the first stage without connection between L0 and L1. In fact, we calculate the single layer grand canonical energies and then use a Legendre transformation to obtain the Helmholtz free energies. The total free energy of the connected system
\begin{align}
F_{\rm{tot}}=F_0(n) +F_1(n_1)+ F_{\rm{es}}(n,n_1)\notag
\end{align}
is the sum of the single layer energies and the electrostatic term $F_{\rm{es}}$. The total energy $F_{\rm{tot}}$ is a function of the total density $Q=n+n_1$ only, that is, the sum of both single layer densities: the dependence on the single layer densities is eliminated by the condition that $n$ has to minimize the total free energy, which is equivalent to the equality of the electrochemical potentials of both layers. Note that, as the layers are connected, $\alpha_{\rm{R}}$ becomes a function of the electric field generated by L1 and thus of $Q$ and $n(Q)$.

Since both layers are coupled through the electric field and are supposed to be in electrical contact, only the inverse compressibility of the total system
\begin{align}\label{kappa}
\kappa^{-1}=Q^2\,\frac{\D^2\left(F/A\right)}{\D Q^2},
\end{align}
is a well-defined thermodynamic quantity. Here, $F=F(Q,V)$ is an appropriate Legendre transform of $F_{\rm{tot}}(Q,n)$ with respect to the interface charge density $n$: the thermodynamic variable is then the voltage bias $V$ between interface and surface (see Appendix~\ref{appCompAndCap}). The electronic compressibility derives from the second derivative of the electronic energy with respect to the volume. In the case of two-dimensional electron systems (as in the considered heterostructure) it is the area instead of the volume. This translates into the second derivative with respect to the electronic areal density $Q$.
The second derivative in Eq.~(\ref{kappa}) is to be taken with respect to the total dependence on $Q$, that is, also the $Q$ dependence of the internal field in the polar film enters. The positivity of $\kappa$ then represents a stability criterion with respect to charge separation in the plane directions.

\begin{figure*}[t]
\centering
\includegraphics[width=0.9\textwidth]{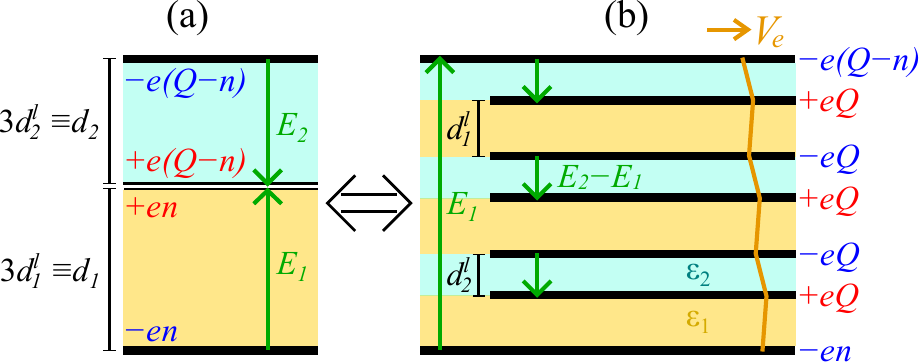}
\caption{(Color online) (a) Multilayered system with static (induced) charges (red/blue), electrical potential perpendicular to the layers (orange) and electric fields (green). (b) Electrostatic equivalent system which relates to the configurational build-up of a polar LAO-film on top of an LAO/STO interface layer.}
\label{Electrostatic}
\end{figure*}

\section{\label{sec:metallicLayer}Polar material with a RSOC interface to a non-polar insulator}

To approach a polar system with a surface and an interface to a non-polar insulator, we first present a model system that consists of three two-dimensional layers (Fig.~\ref{Electrostatic}(a)) but only two of them comprise mobile electrons: The planes at the bottom and at the top have areal charge densities of $-ne$ and $-(Q-n)e$, respectively.  The electrons may be transferred between these two planes to minimize the free energy, while the plane sandwiched between them---with a distance $d_1$ to the bottom and $d_2$ to the top---holds immobile, fixed charges of density $Q$. This charge density $Q$ compensates the charge of the other two planes and therefore keeps the system neutral. The dispersion of the electrons in the bottom layer (L0) is given by Eq.~(\ref{rsocdispersion}), while the top layer (L1) is described as an electron system with a density of states (DOS) of $N_1$. If $n\neq 0$, an electric field perpendicular to L0 exists which derives from the total charge of the two other layers. This field induces the RSOC in the bottom layer and is proportional to the electron density $n$ of L0. The electrostatic energy per area of this layout is
\begin{align}\label{mlotorltotalenergy}
F_{\rm{es}}(Q,n)/A = \frac{2\pi d_1e^2}{\varepsilon_1}n^2+ \frac{2\pi d_2e^2}{\varepsilon_2}\left(Q-n\right)^2,
\end{align}
where $\varepsilon_i$ are the dielectric constants of the material between the planes (see Appendix \ref{appElectrostaticEnergy}). This arrangement is an effective model for the oxide heterostructure LAO/STO, where polar layers with a distance $d^l_1$ and $d^l_2$ between them alternate (Fig.~\ref{Electrostatic}(b)). Due to electronic reconstruction, electrons from the surface relocate to the former empty layer at the LAO/STO interface and occupy $\mathrm{Ti}_{3d}$ states~\cite{Hwang,Thiel}.
We assume that the effective distances to the sandwiched layer are equal, i.e., $d_1/\varepsilon_1=d_2/\varepsilon_2\equiv d/ (2\varepsilon)$. In LAO, the distances $d_1$ and $d_2$ are approximately equal but $\varepsilon_1/\varepsilon_2$ is not accurately known. Therefore we take a layer-independent dielectric constant $\varepsilon$ for the LAO film. This simplification is of no qualitative consequences for the further evaluation; the results depend on an effective distance $d/\varepsilon$.

There are various models for the density dependence of $\alpha_{\rm R}$ (cf.\ Ref.~\cite{Winkler}). In general, the strength of the RSOC $\alpha_{\rm R}$ is taken to be proportional to the SBEF: $\alpha_{\rm R}\propto E$. 
The SBEF $E$ is induced by the charge carrier density $+ne$ (see Fig.~\ref{Electrostatic}(a)) and can be expanded as $E(n) = E_0 +{\cal E}\cdot n + O(n^2)$, where ${\cal E}\cdot n = E_1/2$. 
The factor $1/2$ accounts for the fact that the SBEF is generated by the charge carrier density $+ne$ at the sandwiched layer
and not by the electron density $-ne$ at the LAO/STO interface L0. The field $E_0$ refers to an electric field that is already
present in the absence of a charge density $+ne$; it may be produced by dipolar distortions in the LAO-layer next to L0. With this expansion of $E(n)$, one can also decompose 
\begin{align}\label{alphaR}
\alpha_{\rm{R}}(n)=\alpha_0+ \alpha_1\cdot n +O(n^2).
\end{align}
A more adequate determination of $\alpha_{\rm R}(n)$ will be given in the context of the 6-band model for the LAO/STO interface in Sec.~\ref{sec:3band}.

In order to implement the difference in chemical potentials of the electronic systems at the interface L0 and surface L1,
we introduce a term  $-V_0\,n$ in the total free energy $F_{\rm{tot}}$:
\begin{align}\label{totalenergy}
F_{\rm{tot}}(Q,n)=F_0(n) +F_1(Q-n)+ F_{\rm{es}}(Q,n) -e V_0\,n A.
\end{align}
The potential $V_0$ may be determined from electronic structure evaluations; we use it phenomenologically to tune the electron density at layer L0 to the experimentally observed values.

\begin{figure*}[t]
\centering
\includegraphics[width=0.9\textwidth]{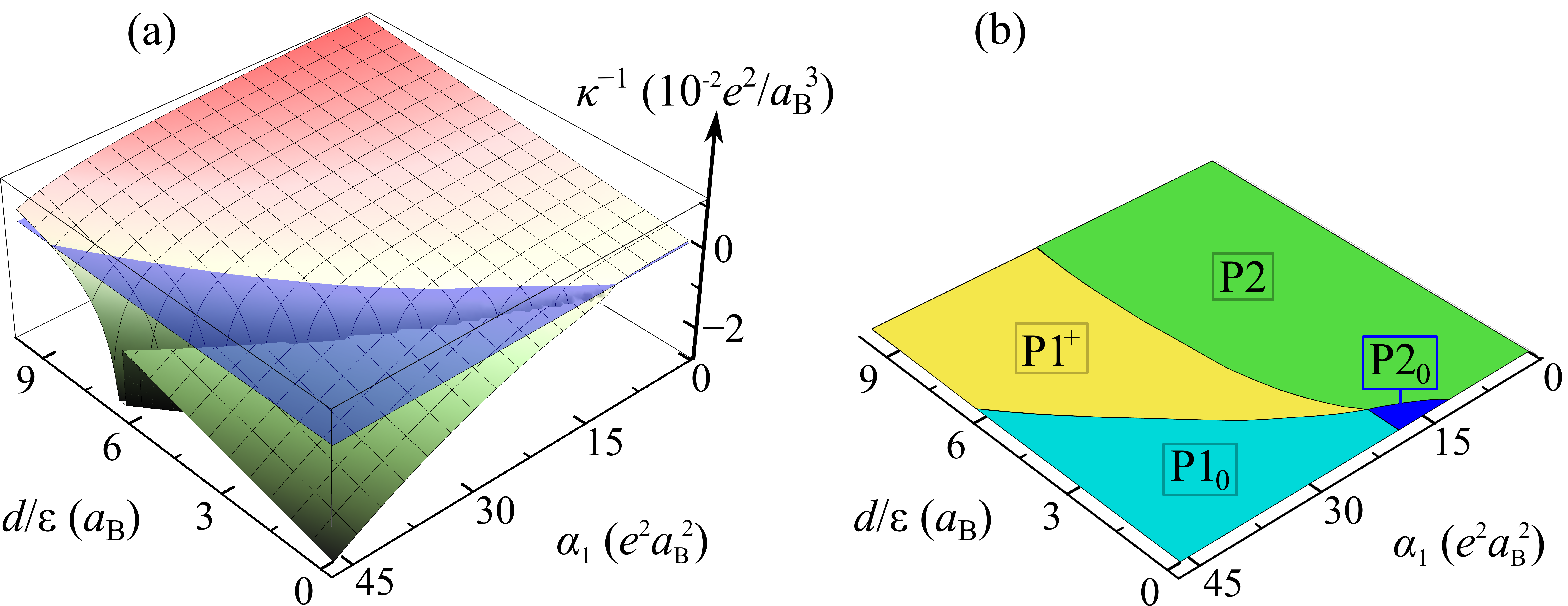}
\caption{(Color online) (a) Inverse compressibility over plate distance $d/\varepsilon$ and strength of RSOC $\alpha= \alpha_0+\alpha_1\,n$. The blue plane indicates $\kappa^{-1}=0$. Here we take $m_0=0.7\,m_e$, $m_1=m_e$, $Q=1\rm{e^-}/{\rm uc}=6.6\cdot 10^{14}$/cm$^2$, $V_0=0$, and $\alpha_0=0$. (b) Phase of the system: P1$^+$ (yellow), P2 (green), P1$_0$ (cyan) and P2$_0$ (blue).}
\label{compressibilityoverDal}
\end{figure*}

The thermodynamic relation  ${\partial F_{\rm{tot}}}/{\partial n}=eA\,V(Q,n)$ 
is used to find the function $n(Q,V)$ (here we still consider $V=0$, and $n(Q,0)\equiv n(Q)$). Since $n(Q)$ is limited to the interval $[0,Q]$, it is possible that the boundary values $n=Q$ or $n=0$ yield the global minimum of $F(Q,V)$. If neither all nor no charge resides at the bottom layer, an analytical expression for the stability of the total system can be derived (Appendix~\ref{appCompAndCap}):
\begin{align}\label{mlotorlgeneralcondition}
\kappa > 0 \Longleftrightarrow \frac{\partial n}{\partial Q} < 1 .
\end{align}
We would like to point out that a system with $\kappa <0$ can still be stabilized by a mechanism not related to the electronic system in L0 and L1 (and thereby yield a positive total compressibility). In this case,  an enhancement of the total electronic charge in L1 causes an amount of charge larger than the added charge to flow to the RSOC layer L0. The distribution of charge between L0 and L1 is controlled by the electric field which is in turn determined by the positive capacitance of the system. 
On the contrary, if the system cannot be stabilized (that is, the total compressibility is negative) then it is expected to develop a phase separated (inhomogeneous) state in the plane. This latter scenario will not be investigated here but we will comment in our conclusions on the proposition~\cite{Grilli,Bucheli,Seibold} of RSOC-driven phase separation.

We will not discuss the temperature dependence of the capacitance and therefore reduce our considerations to zero temperature in the following evaluations.
The free energy per area of the top two-dimensional layer (L1)
\begin{align}\label{mlotorlmetallicenergy}
F_1(Q-n)/A=\frac{1}{2N_1}\left(Q-n\right)^2
\end{align}
depends on the electron density transferred to the bottom layer. Here we assumed an interactionless electron system in L1, characterized by a density of states $N_1$. With $N_1\rightarrow\infty$ (that is, the mass of the charge carriers in L1 increases beyond limits, $m_1\rightarrow\infty$), a localized state can be generated technically. The surface of the LAO/STO heterostructure is insulating. Then, an additional metallic layer on top of the heterostructure produces a capacitor. The top electrode can be a gold or YBa$_2$Cu$_3$O$_7$ (YBCO) film, which connects well with the topmost AlO$_3$ layer (see Appendix~\ref{sec:3Dlayer}).
There, the capacitance increase is slightly smaller than for a 2D electrode due to the finite screening length in the 3D film but the qualitative results are the same as presented in this section for the 2D top electrode.

We re-emphasize that the top layer L1 must be included necessarily in order to render the correct electrostatics which then yields the proper total compressibility and capacitance. However, as long as interaction or spin-orbit effects are negligible in L1, the impact of L1 is determined by two parameters, only: the density of states $N_1$ at the surface electrode and the energetic displacement $-eV_0$ between interface and surface electronic levels. The latter controls trivially the electronic density $n$ at the interface. The former yields the kinetic term in the quantum capacitance and is known to produce a slightly increased effective distance between the electrodes (see, e.g., Refs.~\onlinecite{Mead61,Hebard,Luryi,Buettiker93,Kopp}). Consequently, the materials properties of the surface electrode do not affect the discussed physical nature of the heterostructure qualitatively, except for adjusting the carrier density at the interface.

The energy per area of the layer with RSOC (with $\alpha_{\rm R}$ as density independent constant), calculated in Appendix~\ref{appEnergySingleLayers}, is
\begin{align}\label{mlotorlrashbaenergy}
F_0^\lessgtr(n)/A= \left\{ \begin{array}{cl}\displaystyle \frac{n^3}{6\pi N_0^3\alpha_{\rm R}^2}-\frac{\pi}{2}N_0\alpha_{\rm R}^2n &,\mu<0\\
\displaystyle \frac{n^2}{2N_0}-\pi N_0\alpha_{\rm R}^2n +\frac{\pi^2}{6}N_0^3\alpha_{\rm R}^4 &,\mu>0 \end{array}\right.
\end{align}
When taking the density dependence of $\alpha_{\rm R}$ into account, we expand $\alpha_{\rm R}(n)$ in $n$ (cf.~Eq.~(\ref{alphaR})) and keep terms up to linear order in $n$: 
\begin{align}\label{alpha}
\alpha\equiv\alpha_0+ \alpha_1\cdot n.
\end{align}
One has to distinguish between the case where only the lower band of the RSOC split-bands is occupied, $F_0^<$ (see Fig.~\ref{Dispersion} with $\mu<0$), and the case where both bands are occupied, $F_0^>$ (Fig.~\ref{Dispersion} with $\mu>0$). Since $\mu=0$ corresponds to $n=\pi N_0^2\alpha_{\rm R}^2$, the first case equals either $n<n_-$ or $n>n_+$, with
\begin{align}
n_\pm=\frac{1}{2\pi(N_0\alpha_1)^2}\left(1-2\pi N_0^2\alpha_0\alpha_1\pm\sqrt{1-4\pi N_0^2\alpha_0\alpha_1}\right). 
\end{align}
This means that also for large densities $n>n_+$, all electrons are in the lower band, because the minimum energy of the bands, $F_{\rm{min}}=-\pi N_0(\alpha_0+\alpha_1\, n)^2/2$, decreases for increasing density.\\

With the total energy given by Eqs.~(\ref{mlotorltotalenergy}), (\ref{mlotorlmetallicenergy}) and (\ref{mlotorlrashbaenergy}), we calculate the minimizing density $n(Q)$ numerically, and analytically for $\alpha_0=0$. We identify six different solutions for $n(Q)$ and determine the respective compressibility:\\
\underline{$n=0$} (P0): The bottom layer L0 is empty and the total compressibility is given by the compressibility of the top layer.\\
\underline{$0<n<n_{-}$} (P1$^-$): This phase exists only for $\alpha_0>0$. All electrons of the bottom layer are in the lower band. The compressibility is calculated numerically according to Eq.~(\ref{appComp2}).\\
\underline{$n_{-} <n<\min(Q,n_+ )$} (P2): Both bands of the bottom layer are occupied. The compressibility is calculated numerically according to Eq.~(\ref{appComp2}).\\
\underline{$n=Q<n_+$} (P2$_0$): All electrons reside in the bottom layer, and both bands are occupied. If $n=Q$ is inserted into the total energy $F_{\rm tot}$ and the second derivative with respect to $Q$ is taken, the inverse compressibility is
\begin{align}\label{kappaP20}
\kappa^{-1}_{\rm P2_0}&=\frac{Q^2}{N_0}\bigg(1-2\pi N_0^2\alpha_1(2\alpha_0+3\alpha_1Q)\notag\\
&\qquad+2\pi^2N_0^4\alpha_1^2(\alpha_0+\alpha_1Q)^2+2\pi e^2 N_0 \frac{d}{\varepsilon}\bigg).
\end{align}
\underline{$n_+<n<Q$} (P1$^+$): All electrons of the bottom layer are in the lower band. The compressibility is calculated numerically according to Eq. (\ref{appComp2}).\\
\underline{$n_+<n=Q$} (P1$_0$): All electrons reside in the bottom layer, but only the lower band is occupied. Inserting $n=Q$ into $F_{\rm tot}$ yields for the total compressibility:
\begin{align}\label{kappaP10}
\kappa^{-1}_{\rm P1_0}&=\frac{Q^2}{N_0}\bigg(\frac{\alpha_0^2Q}{\pi N_0^2(\alpha_0+\alpha_1Q)^4}\notag\\
&\qquad-\pi N_0^2\alpha_1(2\alpha_0+3\alpha_1Q)+2\pi e^2 N_0 \frac{d}{\varepsilon}\bigg) .
\end{align}
In Fig.~\ref{compressibilityoverDal}(a) we display the inverse compressibility in the respective phases as a function of the effective plate distance $d/\varepsilon $ and of $\alpha_1$ for $\alpha_0=0$ and a total charge of one electron per areal unit cell of the LAO/STO structure. The transition to a regime of negative compressibility is indicated by the transparent blue plane: the area below the blue plane exhibits negative compressibility, which can occur for all phases except P0. For phase P2, however, the negative compressibility regime is small, restricted to very small $d/\varepsilon$ (less than $\sim 0.2\, a_{\rm B}$) and large RSOC ($\alpha_1 \gtrsim 10 \,(e a_{\rm B})^2 $). For large effective plate distances both layers are always occupied, since the electrostatic term is then dominating. In the reverse case---for small very $d/\varepsilon$---the electrostatic cost for transferring all electrons to one layer is negligible. Then, as $n_+$ decreases with increasing $\alpha_1$, only one band in L0 is occupied for strong RSOC.

\begin{figure*}[t]
\centering
\includegraphics[width=0.9\textwidth]{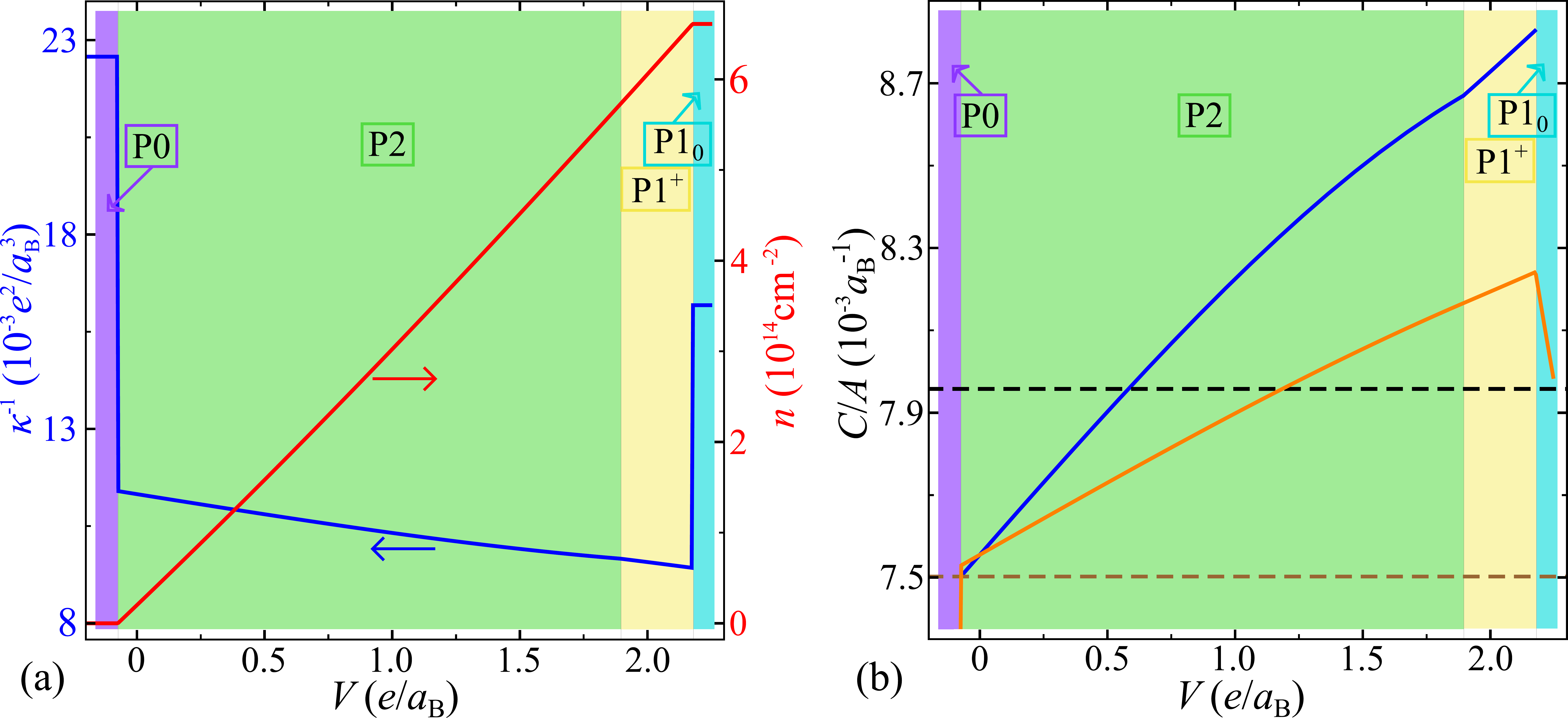}
\caption{(Color online) Inverse compressibility, areal density, and capacitance for a two-band interface model as a function of the bias voltage $V$, for which $e/a_{\rm B}$ corresponds to 27.2~V in SI units. Here $m_0=0.7\,m_e$, $m_1=m_e$, $d/\varepsilon=10\,a_{\rm B}$, $Q=1e^-/{\rm uc}$, $\alpha_0=0$, and $\alpha_1 = 20\,e^2a_{\rm B}^2$ for a two layer system. (a) Inverse compressibility of the total system obtained as analytical solution for the one-band case (blue). Density $n$ on the interface layer (red). (b) Differential capacitance per area calculated analytically (blue), geometric capacitance per area (black dashed), capacitance per area without RSOC (brown dashed) and capacitance $e\Delta n/\Delta V$ per area (orange). The background color indicates the phase of the system according to the caption of Fig.~\ref{compressibilityoverDal}, while violet corresponds to P0.}
\label{nandcompoverV}
\end{figure*}

For $\alpha_0=0$ the phase boundary between P1$^+$ and P1$_0$ is given by 
$$\frac{\pi e^2d}{\varepsilon}=\frac{(3\pi N_0^2\alpha_1^2Q)^2-1}{12\pi\, N_0^3\,\alpha_1^2\,Q}.$$ 
This relation can be used to show analytically that $\kappa_{\rm P1_0}$ is always negative. One deduces from Eq.~(\ref{kappaP20}) for phase P2$_0$ that also $\kappa_{\rm P2_0}<0$. Obviously, the phases P1$_0$ and P2$_0$ display a ``phase separation between the layers'' (the index 0 indicates that all mobile charge carriers are in layer L0). Moreover, the negative compressibilities ($\kappa_{\rm P1_0}<0$, $\kappa_{\rm P2_0}<0$) signal that these phases are thermodynamically unstable and that a phase separation within plane L0 is expected.

When no  charge resides on the surface---as it is the case in phases P1$_0$ and P2$_0$---one could expected that the compressibilities of Eqs.~(\ref{kappaP20}) or (\ref{kappaP10}) are identical to the compressibilities calculated for one isolated layer in Ref.~\cite{Grilli} (apart from the contribution of the electrostatic energy which straightforwardly produces the geometrical inverse capacitance). This is, however, not the case: our results differ from those of Ref.~\cite{Grilli}, as explained in Appendix~\ref{appEnergyVsChemPot}. This discrepancy can be traced back to the inclusion of a $n$-dependent RSOC $\alpha_{\rm R}(n)$ either directly in the free energy or in a later stage in the chemical potential. We calculate the inverse compressibility through the second derivative of the free energy with respect to $Q=n$ whereas the authors of Ref.~\cite{Grilli} identify the inverse compressibility through a first derivative of the chemical potential. The latter approach misses a non negligible contribution from the derivative of the energy with respect to the density in $\alpha_{\rm R}(n)$. For the phase P1$_0$, for example, the latter approach results in an inverse compressibility that is three times smaller than our result for phase P1$_0$. Moreover, those phases which do not have all electrons transferred from the surface L1 to the interface L0, cannot be addressed by the single layer scheme in Ref.~\cite{Grilli}, and we identify them from the free energy which comprises an $n$-dependent RSOC $\alpha_{\rm R}(n)$.

The application of a voltage bias $V$ between interface and surface, i.e., between L0 and L1, can be used to tune the electron density on the interface and hence also the phase of the system (see Fig.~\ref{nandcompoverV}): As $n$ increases, so does the strength of the RSOC $\alpha$, and the band minimum is shifted to lower energies until only one band is partially occupied and the system is in state P1$^+$. This process increases the compressibility that stays positive even if all charge is transfered to the interface (P1$_0$). This does not contradict the analytical analysis from above, which was for $V=0$. 

As for the phases P0 and P1$_0$, note that $\kappa_{\rm P0}$ is independent of $\alpha_0$, $\alpha_1$ and $N_0$, as no charge is in L0, and that $\kappa_{\rm P1_0}$ does not depend on the second electrostatic term in Eq.~(\ref{mlotorltotalenergy}), as no charge is in L1. This observation explains the sharp transitions in $\kappa(V)$ in Fig.~\ref{nandcompoverV}(a).

The differential capacitance $C_{\rm{diff}}$ is derived in appendix \ref{appCompAndCap}:
\begin{align}
A/C_{\rm diff}=\frac{1}{e}\frac{\partial V}{\partial n}=\frac{\partial_{n}^2F_0+\partial_{n}^2F_1+\partial_{n}^2 F_{\rm es}}{A\cdot e^2}.
\end{align}
The term $\partial_{n}^2 F_{\rm{es}}$ yields the geometrical capacitance $C_{\rm geo}=A\varepsilon/4\pi d$. The total capacitance is enhanced above the geometrical value, if the sum of the inverse capacitances of the single layers is negative. In Fig.~\ref{nandcompoverV} we compare $C_{\rm diff}$, the capacitance $C=e\Delta n/\Delta V$, the geometric capacitance $C_{\rm geo}$, and $C_{\alpha=0}=A/(4\pi d/\varepsilon+1/e^2N_0+1/e^2N_1)$, the capacitance for an interface without RSOC. It is evident that the differential capacitance is enhanced above the capacitance of a system without RSOC for finite interface electron densities.
Moreover, both $C_{\rm diff}$ and $C$ increase beyond $C_{\rm geo}$ for sufficiently large $V$ (Fig.~\ref{nandcompoverV}(b)). However, this does not signify a negative compressibility $\kappa$ (cf.Fig.~\ref{nandcompoverV}(a))---although the partial compressibility of the interface my be negative (the latter is not a well-defined thermodynamic quantity).
As elaborated in Appendix~\ref{appCompAndCap}, the connection between the compressibility of the system and its differential capacitance is
\begin{align}
\kappa^{-1}\cdot A=Q^2\frac{\partial^2F}{\partial Q^2}\left(1-\frac{C_{\rm diff}}{e^2A^2}\frac{\partial^2F}{\partial Q^2}\right),
\end{align}
where $F(Q,V)$ is the Legendre transform of $F_{\rm{tot}}(Q,n)$: $F(Q,V) = F_{\rm{tot}}(Q,n(Q,V)) - eA\,V\cdot n(Q,V)$.
With the general assumptions that the total energy $F_{\rm tot}$ of the system is a functional of $n$ and $Q-n$ exclusively, and that $n$ is not a boundary value ($Q$ or 0) then the compressibility of the system is related to the differential capacitance through
\begin{align}\label{CompressibilityCapacitance}
\frac{1}{Q^2}\,\kappa^{-1}=\frac{e^2A}{C_{\rm diff}}\left(1-\frac{\partial n}{\partial Q}\right)\frac{\partial n}{\partial Q}. 
\end{align}
where both, $\kappa$ and $C_{\rm diff}$, are functions of $Q$ and $V$.
The first factor on the rhs is what one should expect for a single metallic electrode: the electronic compressibility is the capacitance of the electrode up to a factor $e^2Q^2A$. The further two factors with the derivative 
${\partial n}/{\partial Q}$ account for the charge redistribution between surface and interface---which, of course, depends implicitly on the RSOC (see Appendix~\ref{appCompAndCap}).

\begin{figure*}[t]
\centering
\includegraphics[width=0.9\textwidth]{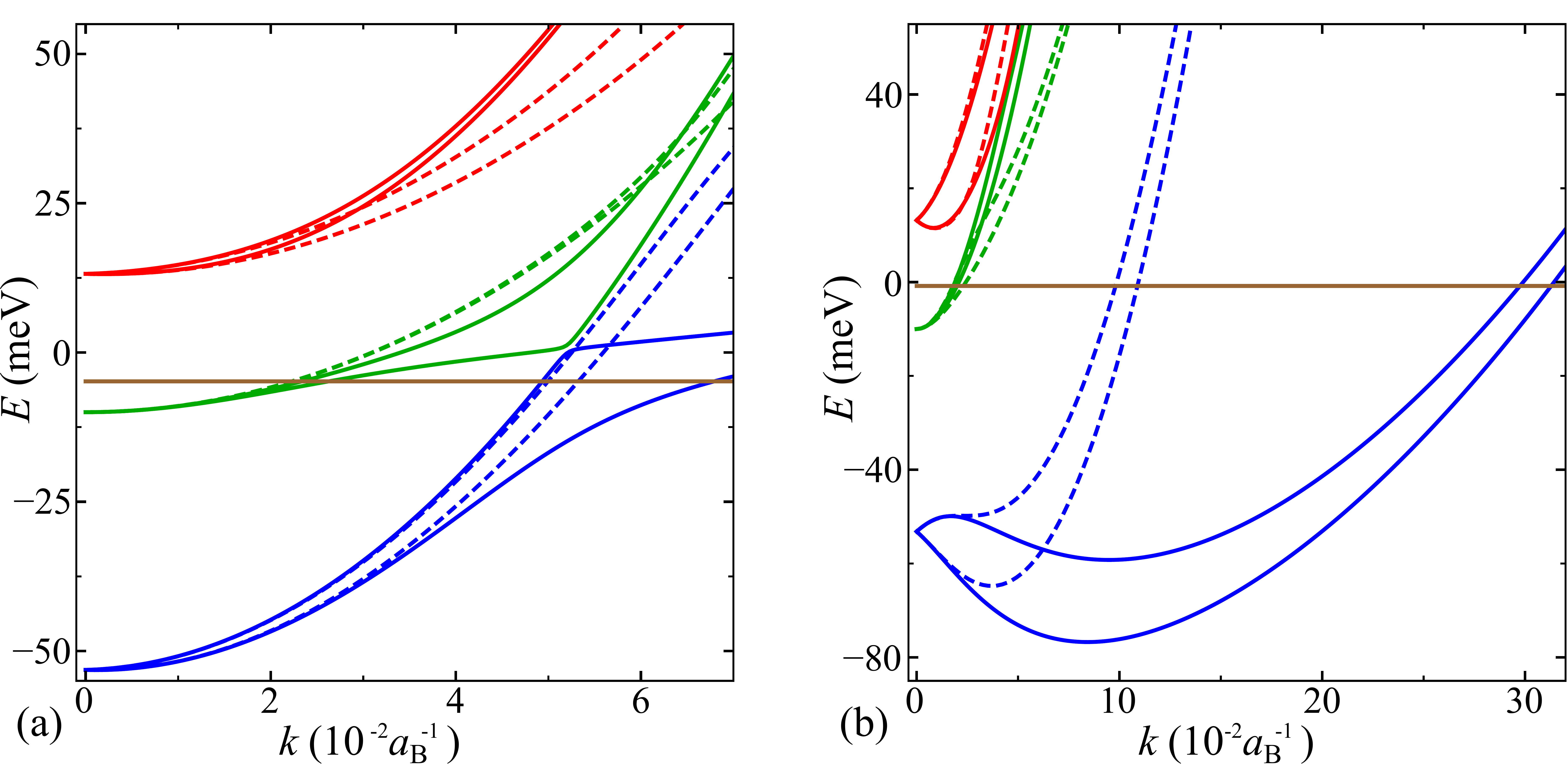}
\caption{(Color online) Energy bands for $m_{\ell}=0.7\,m_{\rm e}$, $m_{\rm h}=15\,m_{\rm e}$, $\Delta_{\rm SO}=10\,{\rm meV}$, $\Delta_{\rm E}=50\,{\rm meV}$, $\Delta_{\rm z,0}=0.61\cdot 10^{-3}\,e^2$, and $\Delta_{\rm z,1}=8.29\,e^2a_{\rm B}^2$. The bands are plotted from the $\Gamma$-Point to the $X$-point (continuous curves) and to the $M$-point (dashed curves). The highest occupied energy level for the densities $n=2\cdot10^{13}$/cm$^2$ (a) and $n=2\cdot10^{14}$/cm$^2$ (b) is indicated by the brown curve.}
\label{MultibandBands}
\end{figure*}

Finally, we discuss the size of the experimental values of the control parameters. 
We use $m_0=0.7\,m_e$ for the effective mass of the electrons at the LAO/STO interface. This value results from the multi-band evaluation~\cite{Joshua}, addressed in the next section: a light mass $m_{\ell} = 0.7\,m_e$ and a heavy mass $m_{\rm h}=15\,m_e$ were identified from ARPES measurements~\cite{Santander-Syro} for the electrons of the $d_{\rm xy}$-, $d_{\rm xz}$- and $d_{\rm yz}$-bands. Joshua \textit{et al.} illustrated in the Suppl.\ of Ref.~\cite{Joshua} that, within the multiband model, the electrons of the lowest resulting band acquire an effective mass of $m_0=0.7\,m_e$ close to the band minimum.

Estimating $\alpha_0$ and $\alpha_1$ has the largest uncertainty: Caviglia~\textit{et al.}~\cite{Caviglia} concluded from their magnetotransport experiments that $\alpha_{\rm R}^{\rm ex}\approx (1\dots 5)\times 10^{-12}\,{\rm eVm}\approx (0.74\dots 3.7)\times 10^{-3}\,e^2$ in the range of backgate voltages where the superconducting dome is formed. For a more negative bias the interface becomes depleted and the RSOC is nearly constant with a low value of $\alpha_0\simeq 10^{-12}\,{\rm eVm}\approx 0.74\times 10^{-3}\,e^2$. We neglect this rather small constant contribution with respect to the term linear in the bias. The RSOC increases approximately linearly with bias in between values corresponding to the low-density foot and top of the superconducting dome (compare Fig.~3 in Ref.~\cite{Caviglia}). The difference in charge carrier density at foot and top of the dome~\cite{Richter} is $\approx 0.015\,e^-/{\rm uc}=9.9\times 10^{12}{\rm /cm}^{2}$, so that
\begin{align}
\alpha_1 &\approx\frac{4\cdot 10^{-12}\,{\rm eVm}}{9.9\cdot10^{12}\,{\rm cm}^{-2}}\approx 4.0\cdot 10^{-23}\,{\rm eV\cdot cm}^{3}\notag\\
&\approx 10\,(ea_{\rm B})^2.
\label{approxalpha1}
\end{align}
We do not know the exact values for the density at which $\alpha_{\rm R}^{\rm ex}$ was measured. Therefore $\alpha_1\approx 10\,(ea_{\rm B})^2$, as deduced from the experiment of Caviglia~\textit{et al.}~\cite{Caviglia}, might be somewhat smaller but certainly not more than one order of magnitude.

\section{Multiband Layer with Spin-Orbit Coupling}
\label{sec:3band}

\begin{figure*}[t]
\centering
\includegraphics[width=0.9\textwidth]{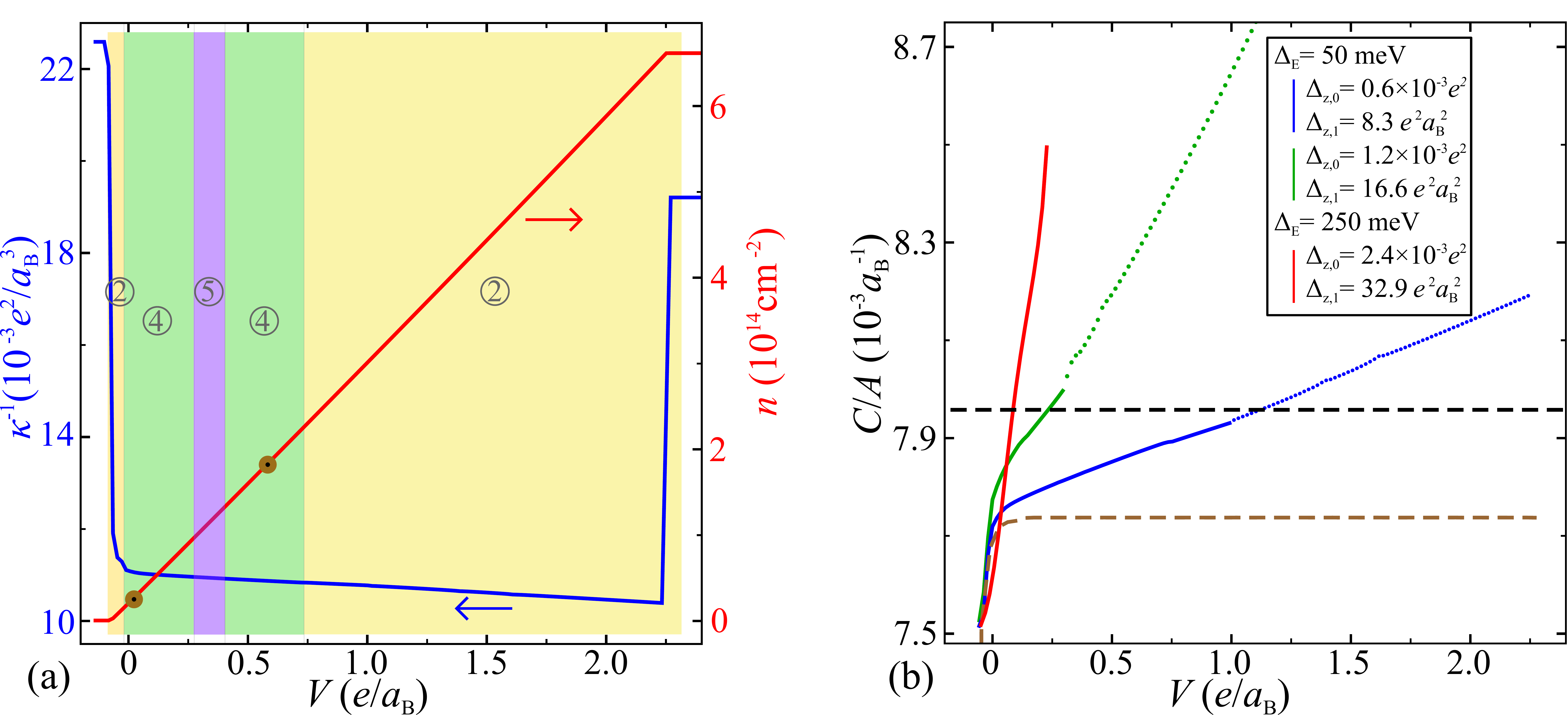}
\caption{(Color online) Inverse compressibility, areal density, and capacitance for a six-band interface model as a function of the bias voltage $V$. 
Here $m_{\ell}=0.7\,m_{\rm e}$, $m_{\rm h}=15\,m_{\rm e}$, $m_1=\,m_{\rm e}$, $d/\varepsilon=10\,a_{\rm B}$, $Q=1e^-/{\rm uc}$, and $\Delta_{\rm SO}=10\,{\rm m}e{\rm V}$, and $e/a_{\rm B}$ corresponds to 27.2~V in SI units. (a) Inverse compressibility of the total system (blue) and density on the interface (red) for $\Delta_{\rm z,0}=0.61\cdot 10^{-3}\,e^2$, $\Delta_{\rm z,1}=8.29\,e^2a_{\rm B}^2$, $\Delta_{\rm E}=50\,{\rm meV}$, and $V_0=-1.149\,e/a_{\rm B}$. The background colors indicate the number of occupied bands: two (yellow), four (green) and five (purple). The brown circles refer to systems with bias values $V$ that correspond to the densities $n$ of Figs.~\ref{MultibandBands}(a) and (b), respectively. (b) Differential capacitance per area for three different sets of parameters with $n=2\cdot10^{13}$/cm$^2$ at $V=0$, which is provided by $V_0=-1.149\,e/a_{\rm B}$ (blue), $V_0=-1.149\,e/a_{\rm B}$ (green) and $V_0=-1.155\,e/a_{\rm B}$ (red). For small ${\bf k}$, the energy splitting $\Delta E$ (as in Fig.~\ref{MultibandSplitting}) for the blue and red curve corresponds to $\alpha_1=2.5\,e^2a_{\rm B}^2$ and for the green curve to $\alpha_1=5.0\,e^2a_{\rm B}^2$. For the dotted part of the curves, a $k$-space region around the ${\bf k}={\bf 0}$ wave vector is unoccupied. The geometric capacitance per area (black) and the differential capacitance per area for $V_0=-0.781\,e/a_{\rm B}$, $\Delta_{\rm z,0}=0.3\cdot 10^{-3}\,e^2$ and $\Delta_{\rm z,1}=0$ (brown) are also shown.}
\label{MultibandnCompCap}
\end{figure*}

The free electrons at the LAO/STO interface reside predominantly in the titanium $3d$ orbitals of the first TiO layer.
For a refined description of the interface we use a six-band model for the three spin-split $t_{2g}$ orbitals~\cite{Joshua,Zhong,comment2}. Within this formalism, the emergence of a strong Rashba-like SOC is explained naturally through the breaking of inversion symmetry at the interface, which couples the atomic SOC of the Ti atoms to the inter-band hopping. The Hamiltonian can be written in the $d_{\rm yz}$, $d_{\rm zx}$, $d_{\rm xy}$ basis as:
\begin{align}
\mathcal{H}= \mathcal{H}_{\rm k}+\mathcal{H}_{\rm SO}+\mathcal{H}_{\rm z}.
\end{align}
The first term is the intra-orbital hopping:
\begin{align}
\mathcal{H}_{\rm k}=\left(\begin{matrix}\frac{k_{\rm x}^2}{2\pi N_{\rm h}}+\frac{k_{\rm y}^2}{2\pi N_{\rm l}}&0&0\\	0&\frac{k_{\rm x}^2}{2\pi N_{\rm l}}+\frac{k_{\rm y}^2}{2\pi N_{\rm h}}&0\\	0&0&\frac{k_{\rm x}^2+k_{\rm y}^2}{2\pi N_{\rm l}}-\Delta_{\rm E}\end{matrix}\right)\otimes\sigma_0,
\end{align}
$N_{\rm i}=m_{\rm i}/(\pi\hbar^2)$ is the density of states for light ($m_{\ell}$) or heavy ($m_{\rm h}$) electrons and $\Delta_{\rm E}$ the splitting between the $d_{\rm xy}$- and the $d_{\rm yz}$-, $d_{\rm zx}$-orbitals. The contribution from atomic spin-orbit coupling is
\begin{align}
\mathcal{H}_{\rm SO}=\Delta_{\rm SO}\cdot\sum_{i=x,y,z}L_i\otimes\sigma_i
\end{align} 
\begin{align}
{\bf{L}}=\left( \left(\begin{matrix}0&0&0\\0&0&i\\0&-i&0\end{matrix}\right),\left(\begin{matrix}0&0&-i\\0&0&0\\i&0&0\end{matrix}\right),\left(\begin{matrix}0&i&0\\-i&0&0\\0&0&0\end{matrix}\right)\right)\hbar
\end{align}
where $\Delta_{\rm SO}$ is the strength of the atomic SOC,  $\sigma_0$ is the 2$\times$2 identity matrix, and $\sigma_{i}$ are the 2$\times$2 Pauli-matrices. Due to the confining potential at the interface and its asymmetry, the $d$-orbitals are deformed and hence no longer orthogonal. This allows (anti-symmetric) interorbital hopping between $d_{\rm xy}$- and $d_{\rm xz}$-orbitals along the $y$-direction and between $d_{\rm xy}$- and $d_{\rm yz}$-orbitals along the $x$-direction, presented by
\begin{align}
\mathcal{H}_{\rm z}=\Delta_{\rm z}\left(\begin{matrix}0&0&-ik_{\rm x}\\	0&0&ik_{\rm y}\\	ik_{\rm x}&-ik_{\rm y}&0\end{matrix}\right)\otimes\sigma_0.
\end{align}
We assume that the magnitude of the hopping is proportional to the asymmetry inducing electrical field. In our modeling, the origin of this electric field is the (positive) charge on the surface that compensates for the electrons transferred from the surface to the interface system. Therefore we write: 
\begin{align}
\Delta_{\rm z}=\Delta_{\rm z,0}+\Delta_{\rm z,1} \cdot n.
\end{align}
The term proportional to $n$ is generated through the asymmetry inducing electrical field, and the $n$-independent term 
$\Delta_{\rm z,0}$ can be traced to dipolar distortions in the LAO-layer next to the interface.
The diagonalization, which can be done analytically only for $k_{\rm x}=k_{\rm y}=0$, yields three spin split doublets. $\mathcal{H}_{\rm z}$, which originates from the deformation, vanishes at ${\bf k}={\bf 0}$ and the doubly degenerate energies are
\begin{subequations}
\begin{align}
\epsilon_1(k_{\rm x}=0,k_{\rm y}=0)&= \frac{1}{2}\left(-\Delta_{\rm E}+\Delta_{\rm SO}-S\right),\\
\epsilon_2(k_{\rm x}=0,k_{\rm y}=0)&= -\Delta_{\rm SO},\\
\epsilon_3(k_{\rm x}=0,k_{\rm y}=0)&= \frac{1}{2}\left(-\Delta_{\rm E}+\Delta_{\rm SO}+S\right)\\
{\rm with}\;S&=\sqrt{\Delta_{\rm E}^2+2\Delta_{\rm E}\Delta_{\rm SO}+9\Delta_{\rm SO}^2}.\nonumber
\end{align}
\end{subequations}
For a non-zero $\Delta_{\rm z}$, the degeneracies are lifted. Moreover, crossings between different doublets are avoided~\cite{Joshua,Zhong}  (see Fig.~\ref{MultibandBands}a). With increasing $\Delta_{\rm z}$, the splitting between the formerly degenerated bands becomes larger and the point of lowest energy in $k$-space moves to larger values of 
$k$ (cf.~Fig.~\ref{MultibandBands}(b)). 

\begin{figure}[b]
\centering
\includegraphics[width=0.9\columnwidth]{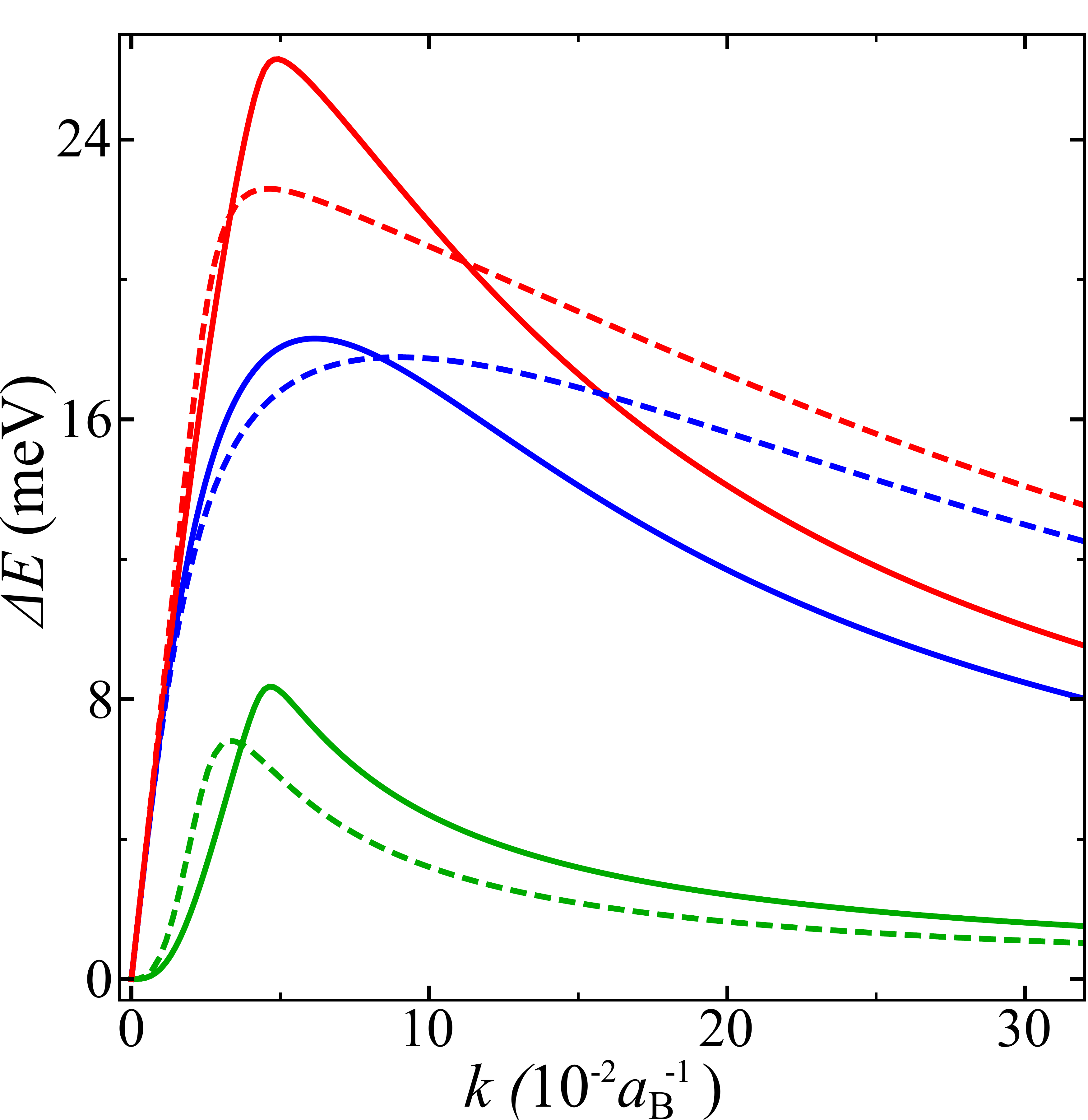}
\caption{(Color online) Energy splitting between the two lowest bands (blue), the third and fourth band (green) and the two highest bands (red). Here the parameters are $m_{\ell}=0.7\,m_{\rm e}$, $m_{\rm h}=15\,m_{\rm e}$, $\Delta_{\rm SO}=10\,{\rm meV}$, $\Delta_{\rm E}=50\,{\rm meV}$, $\Delta_{\rm z,0}=0.61\cdot 10^{-3}\,e^2$, $\Delta_{\rm z,1}=8.29\,e^2a_{\rm B}^2$, and $n=2\cdot 10^{14}$/cm$^2$, as for the right side of Fig. \ref{MultibandBands}. The energy splitting is plotted from the $\Gamma$-Point to the $X$-point (continuous curves) and to the $M$-point (dashed curves).}
\label{MultibandSplitting}
\end{figure}

The shape of the bands depends, analogously to the two-band model with RSOC (cf.~Eq.~(\ref{rsocdispersion})), on the electron density. We add a top gate as in Sec.~\ref{sec:metallicLayer} and perform the same minimization of the energy of the combined system with respect to $n$ for different applied voltages. The results are shown in Fig.~\ref{MultibandnCompCap}, where we chose the light and heavy electron masses $m_{\ell}=0.7\,m_{\rm e}$ and $m_{\rm h}=15\,m_{\rm e}$, respectively (in accordance with the ARPES~\cite{Santander-Syro} measurements for ${\rm SrTiO_3}$ surfaces). The energy splitting is taken to be $\Delta_{\rm E}=50\,{\rm meV}$, according to the x-ray measurements in Ref.~\cite{Salluzzo}. Alternatively, we also consider $\Delta_{\rm E}=150$\,meV, which is in better agreement with various ab initio evaluations (see, e.g., Ref.~\cite{Zhong}). The atomic SOC was estimated from transport experiments~\cite{Caviglia,Shalom,Joshua} and ab-initio calculations~\cite{Mattheiss} to be $\Delta_{\rm SO}=10$--25\,meV; we chose the lower boundary of 10\,meV.

\begin{figure*}[t]
\centering
\includegraphics[width=0.9\textwidth]{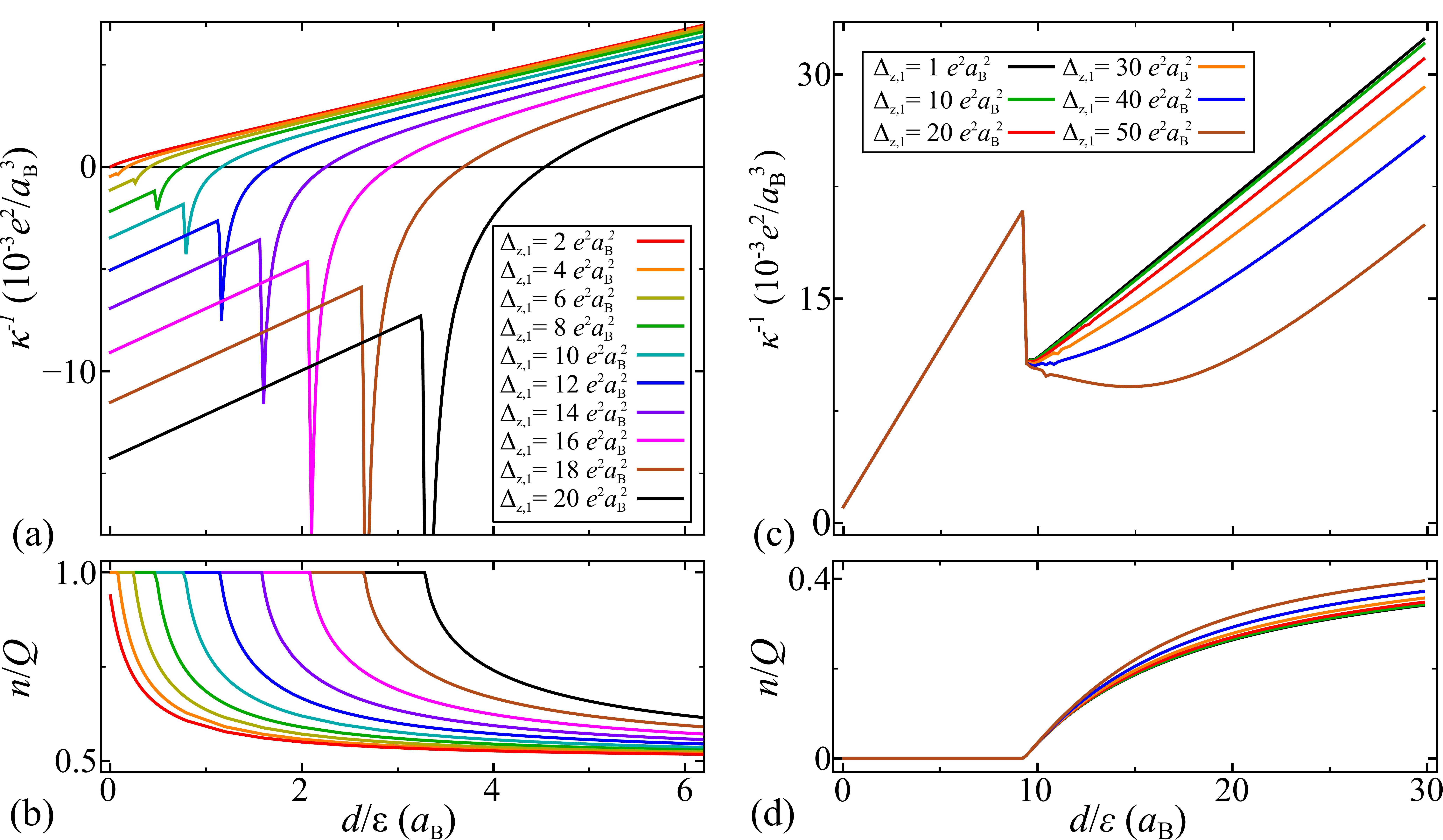}
\caption{(Color online) Inverse compressibility for $V_0=0$ (a) and for $V_0=-31\, {\rm V}$ (c), both as functions of the effective interface-surface distance $d/\varepsilon$. Here $m_{\ell}=0.7\,m_{\rm e}$, $m_{\rm h}=15\,m_{\rm e}$, $m_1=m_{\rm e}$, $Q=1e^-/{\rm uc}$, $\Delta_{\rm SO}=10\,{\rm meV}$, $\Delta_{\rm E}=50\,{\rm meV}$, and $\Delta_{\rm z,0}=10^{-3}\,e^2$. The plots below, i.e., panels (b) and (d) display the interface electronic density $n$ for the respective potentials $V_0$.}
\label{MultibandComp}
\end{figure*}

The $k$-dependent energy splitting $\Delta E(k)$ induced by $\Delta_{\rm z}$ in the lower and upper doublet increases linearly for small $k$, i.e., it is Rashba-like: $\Delta E(k)=\alpha(\Delta_{\rm z})k$ (cf.~Fig.~\ref{MultibandSplitting}). This can be employed to fit the parameters $\Delta_{\rm z,0}$ and $\Delta_{\rm z,1}$ to the estimates of $\alpha_0$ and $\alpha_1$ made from experimental results in the previous chapter: 
by writing
\begin{align}
\alpha(n)=\alpha(\Delta_{\rm z}(n))=b(\Delta_{\rm z,0}+\Delta_{\rm z,1}\cdot n),
\label{approxdelta}
\end{align}
we identify $b\Delta_{\rm z,0}\equiv\alpha_0$ and $b\Delta_{\rm z,1}\equiv\alpha_1$, where $b$ is obtained from fitting $\alpha(\Delta_{\rm z})$ to the numerically calculated $\Delta E(k)$. For $\Delta_{\rm E} = 50\,{\rm meV}$ and $\Delta_{\rm SO}=10\,{\rm meV}$ this yields $b\approx3.3$ and we find $\Delta_{\rm z,0}\approx 2.5\cdot 10^{-3} e^2$ and $\Delta_{\rm z,1}\approx 33\,e^2a_{\rm B}^2$.
Although this estimation for $\Delta_{\rm z,0}$ and $\Delta_{\rm z,1}$ neglects the dependence of $\Delta E$ on the angular direction in $k$-space, it gives a reasonable approximation for $\alpha(n)$ as long as the charge density is small, such that $\Delta E(k)\propto k$ holds. For larger densities, for which this proportionality fails around the Fermi momentum (i.e., it is not Rashba-like, cf. Fig.~\ref{MultibandSplitting}), the above estimation is no longer valid.
The capacitance and compressibility of the system is given mostly by the electrons from the lowest doublet, because for low densities the higher bands are unoccupied or considerably less occupied than the lowest two bands. For larger densities, on the other hand, the variation of the electron number upon changing $V$ or $Q$ is largest in the lowest doublet (since then the DOS is larger in the lower doublet, cf.~Fig.~\ref{MultibandBands}(b)). 

The differential capacitance $C_{\rm diff}$ for different gate voltages $V$ is displayed in Fig.~\ref{MultibandnCompCap}(b). The difference of the chemical potentials in the interface and the metallic top layer, $V_0$, was set to a value that gives a charge density $n=2\times 10^{13}/{\rm cm}^2$ for $V=0$. For $\Delta_{\rm z,1}=0$, i.e., a density independent bandstructure, the capacitance increases with voltage (see brown line in Fig.~\ref{MultibandnCompCap}(b)) until the Fermi energy reaches a value where the deviation from a parabolic band structure becomes negligible and, consequently, the DOS is constant. Notice that the capacitance derived from the microscopic model is smaller than the geometric capacitance, since the kinetic energy of the charge carriers in the electrodes adds a positive term to the inverse capacitance (see, e.g., Ref.~\cite{Kopp}).

A non-zero $\Delta_{\rm z,1}$ enhances $C_{\rm diff}$, even above the geometrical capacitance if $\Delta_{\rm z,1}$ is sufficiently large. With increasing voltage $V$, the density $n$ at the interface increases (Fig.~\ref{MultibandnCompCap}(a)), and hence $\Delta_{\rm z}$.
This flattens the dispersion of the lower doublet and thereby enhances the effective mass at $k_{\rm F}$ (see Fig.~\ref{MultibandBands}), which causes the increase of capacitance above the value for $\Delta_{\rm z,1}=0$ (cf.~Fig~\ref{MultibandnCompCap}(b)).  For large $\Delta_{\rm z}$, the energy minima of the two lowest bands are far from the zero point and ${\bf k}={\bf 0}$ is not occupied. In this regime, $\Delta E(k)$ is no longer linear in $k$ and the estimations for $\Delta_{\rm z,0}$ and $\Delta_{\rm z,1}$ are no longer valid. This region is indicated by the dots of the differential capacitance in Fig.~\ref{MultibandnCompCap}(b).

Within the regime where the bands become flatter with increasing density, negative compressibility may appear.
The inverse compressibility for different values of $\Delta_{\rm z,1}$ is shown in Fig.~\ref{MultibandComp} as a function of distance between the interface and surface. Evidently,  Fig.~\ref{MultibandComp}(a), which presents the inverse compressibility for the 6-band model corresponds to  Fig.~\ref{compressibilityoverDal}(a) for the 2-band model. For a small effective distance $d/\varepsilon$, all charge resides on the interface (Fig.~\ref{MultibandComp}(b)), the compressibility is negative, and $\kappa^{-1}$ increases nearly linearly with the distance. This is understood by setting $n=Q$ in Eq.~(\ref{mlotorltotalenergy})---the only term in the total energy with a dependence on $d/\varepsilon$---and taking the total second derivative with respect to $Q$. For large distance the electrostatic energy dominates and the total energy is minimized by $n\approx Q/2$. Inserting this into Eq.~(\ref{mlotorltotalenergy}) yields half the gradient of the $n=Q$ regime, see Fig.~\ref{MultibandComp}(a).

In Fig.~\ref{MultibandComp}(a) the inverse compressibility is displayed for $V_0=0$, whereas in Fig.~\ref{MultibandComp}(c) the potential is $V_0=-31\, {\rm eV}$. For $V_0=0$ and $\Delta_{\rm SO}=0$, the lower band edge at the interface is shifted by $-\Delta_{\rm E}=-50\,{\rm meV}$ with respect to the band edge of the surface electrode. Then, the electrons tend to go to the interface if no other effect such as the RSOC interferes, provided that the electrostatic energy is small---and the compressibility can be negative for sufficiently large $\Delta_{\rm z,1}$. 

However, the behavior of $\kappa^{-1}$ as a function of $d/\varepsilon$ changes drastically when we adjust $V_0$ to $-31\,{\rm V}$ so that $n=2\times 10^{13}/{\rm cm}^2$ for $d/\varepsilon=10\,a_{\rm B}$ and $\Delta_{\rm z,1}=33\,e^2a_{\rm B}^2$. Then $n$ is in the range of  carrier densities established for the LAO/STO interface with $d/\varepsilon\simeq10\,a_{\rm B}$. The inverse compressibility first increases linearly with $d/\varepsilon$ because $n=0$ in the regime of small distances (see Fig.~\ref{MultibandComp}(c)). There is a small offset due to the finite compressibility of the surface electronic state. Then, with a further rise in $d/\varepsilon$, the polar-catastrophe mechanism induces a finite electron density at the interface (Fig.~\ref{MultibandComp}(d)), and the inverse compressibility decreases sharply (Fig.~\ref{MultibandComp}(c)) because a small but finite $n$ reduces the electrostatic term linearly with $n$ (see Eq.~(\ref{mlotorltotalenergy})). Eventually, $\kappa^{-1}$ increases towards its limiting linear behavior for large $d/\varepsilon$ as in  Fig.~\ref{MultibandComp}(a). This analysis allows us to conclude that the compressibility stays strictly positive even for large values of the RSOC if the system parameters are adjusted within a range that is consistent with parameter values previously identified for the LAO/STO heterostructure.

For a thorough determination of the compressibility of the LAO/STO heterostructure additional effects may be relevant: these include oxygen vacancies, their distribution between the planes as well as the lateral extension of the two-dimensional electron liquid at the interface. However, these effects influence primarily the charge carrier density. As the charge carrier density is adjusted to the experimental values through a bias $V_0$ in our analysis, we expect that the evaluation of the compressibility remains valid.

\section{Conclusions}
\label{conclusions}
Low-dimensional charge states at interfaces of complex oxides have been explored intensively in recent years, and it is a combination of scientific and technological interest which drives the research in this field. This is also valid for the work presented in this article: we explored the electronic charge redistribution in a heterostructure of a polar and a non-polar insulator, the paradigmatic realization of which has become the LAO/STO interface of a polar LAO film on the non-polar substrate STO. Understanding the quantum phenomena that control the electronic reconstruction at the interface, and thereby also the charge redistribution between surface and interface, may provide a possible strategy to control capacitances by the electronic properties of the conducting electrodes---as opposed to the standard search for dielectrics with high dielectric constant~\cite{Schlom08}.

Sizeable capacitance enhancements and negative electronic compressibility have been observed for LAO/STO~\cite{Li,Tinkl}. Recently,
these findings were related to a strong RSOC~\cite{Grilli,Bucheli,Seibold}. The RSOC is generated by an electric field that in turn is controlled by the electronic charge density $-en$ at the interface. In fact, a band structure, which depends on the density of the band electrons themselves, allows for a decreasing chemical potential with increasing charge carrier density---assuming that the RSOC strength is sufficiently large. Such a dependence of the chemical potential on the carrier density represents a negative compressibility. Thus, instead of the standard explanation for a negative compressibility in low density semiconductor systems~\cite{Eisenstein} or in oxide heterostructures~\cite{Kopp,Li} through exchange and correlation effects, it may be that spin-orbit effects play an essential role. 

However, in order to settle the realization of a negative compressibility and the possible formation of an inhomogeneous charge-separated state, one has to investigate at least the compressibility of the complete electronic system (in the case that all electronic subsystems are coupled). The compressibility of other subsystems which are not comprised of mobile electrons may also contribute and keep the total compressibility positive. In this article we not only included the mobile interface electrons, but also the surface charges with their respective field. This is to be considered as a minimal setup for two reasons: (1) the RSOC is controlled by the electric field of the charges at the surface, and (2) this electric field and the charge carriers of the surface electrode contribute to the compressibility of the interface-surface system.

A remarkable relation, which only involves the capacitance of the interface itself, $C_0$, and the geometric capacitance $C_{\rm geom}= \varepsilon A/4\pi d$ of the interface-surface system, allows to determine if this system attains a negative compressibility:
\begin{align}\label{negcomp1}
\kappa <0\quad\Longleftrightarrow\quad C_0^{-1}+ 2\,\frac{\pi d}{A\,\varepsilon} < 0,
\end{align}
where $C_0^{-1}=(1/eA)^2\,\partial^2 F_0/\partial n^2$ is found from the second derivative of the interface free energy $F_0$ with respect to the interface electronic density $n$. This relation was derived in Appendix~\ref{appCompAndCap} under the conditions that the interface-surface capacitance $C_{\rm diff}^{-1}=(1/eA)^2\,\partial^2 F/\partial n^2$ is positive and that $n$ does not assume a boundary value (that is, neither $n=0$ nor $n=Q$ holds). Here, $F$ is the free energy of the entire heterostructure. The relation (\ref{negcomp1}) is remarkable in two respects: first, the compressibility $\kappa$ can become negative even when the capacitance $C_{\rm diff}$ is positive, and second, this stability criterion does not depend explicitly on the compressibility of the surface---even though the total inverse compressibility $\kappa^{-1}$ contains a contribution from the surface electron system. Both statements can be traced back to the special interface-surface setup in which the capacitance is not equal to the total compressibility (see Eq.~(\ref{CompressibilityCapacitance})).

The relation (\ref{negcomp1}) can be cast into a more intuitive form (Appendix~\ref{appCompAndCap}):
\begin{align}
\kappa > 0 \Longleftrightarrow \frac{\partial n}{\partial Q} < 1.
\nonumber
\end{align}
This inequality obviously expresses the fact that the system is stable if an additional overall electronic charge $\delta Q$ does not induce a charge transfer $\delta n$ larger than the added overall charge. The compressibility is negative and the electronic state may charge-separate only if $\partial n/\partial Q>1$.

In order to be explicit, we considered first a 2-band model for the interface electronic system (Secs.~\ref{sec:metallicLayer} and Appendix~\ref{sec:3Dlayer}). The bands are split by a RSOC that depends on the charge density of the surface electronic system. The charge densities at interface and surface were calculated from the minimization of the total free energy. This simple model allows to identify a series of electronic phases that depend on the occupation of the interface split bands. The second derivatives of the free energy with respect to the total electronic density $Q$ and the interface electronic density $n$ yield (phase-dependent) compressibility and capacitance, respectively. A negative compressibility arises only for very small effective distances $d/\varepsilon$ between interface and surface, and strong RSOC (see Fig.~\ref{compressibilityoverDal}(a); for example for an RSOC-parameter $\alpha_1 \sim 10 \,(e a_{\rm B})^2$ the effective distance has to be $d/\varepsilon\lesssim 0.2\, a_{\rm B}$ ). Even for RSOC parameters twice as large as the largest RSOC estimated from magneto transport measurements for LAO/STO, the compressibility stays positive (cf.~Fig.~\ref{nandcompoverV}(a)), also for large but sensible values of a bias voltage $V$. This differs from the findings of Refs.~\cite{Grilli,Bucheli}. The discrepancy originates most prominently from the electrostatic term $2\,\pi d/A\,\varepsilon$ in relation (\ref{negcomp1}), which is entirely absent in Refs.~\cite{Grilli,Bucheli}. The more complete treatment of our work yields a compressibility that is positive for the parameter set of the LAO/STO heterostructure and therefore retains the stability of the electronic interface-surface system. 

There is also a more technical difference: here, the compressibility is found from the second derivative of the free energy (see relation (\ref{kappa})). In Refs.~\cite{Grilli,Bucheli}, however, the compressibility is calculated from the derivative of the chemical potential, which was identified from a model with fixed, density-independent RSOC. Only for the derivative of the chemical potential, which yields $\kappa^{-1}$, the RSOC was generalized to include the density dependence. Evidently, the two approaches lead to diverse results even when the analysis is restricted to a single layer model without surface term (see Appendix~\ref{appEnergyVsChemPot}).

\begin{figure*}[t]
\centering
\includegraphics[width=0.9\textwidth]{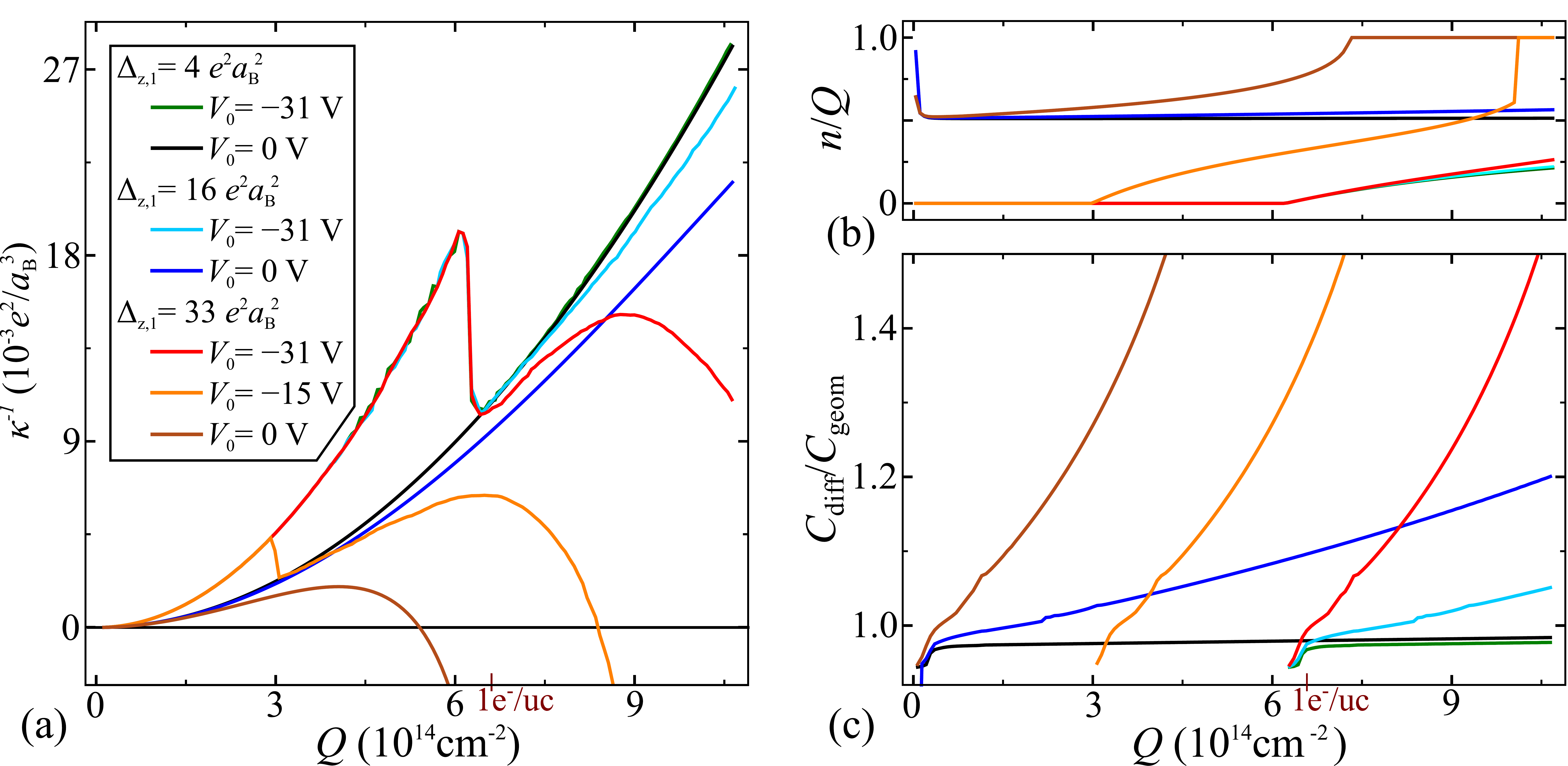}
\caption{(Color online) Compressibility $\kappa$ (a), interface electronic density $n$ (b), and capacitance $C_{\rm diff}$ (c), each in dependence on the polar charge density $Q$.
Here $m_{\ell}=0.7\,m_{\rm e}$, $m_{\rm h}=15\,m_{\rm e}$, $m_1=m_{\rm e}$, $d/\varepsilon=10\,a_{\rm B}$, $\Delta_{\rm SO}=10\,{\rm meV}$, $\Delta_{\rm E}=50\,{\rm meV}$ and $\Delta_{\rm z,0}=10^{-3}\,e^2$. The parameter $\Delta_{\rm z,1}$ (for the hybridization between $d_{\rm xy}$-  and $d_{\rm xz,yz}$-orbitals) controls the density dependence of the RSOC. The voltage $V_0$ allows to tune the interface charge density. 
In order to attain a charge density compatible with experiments one has to choose a negative voltage; the lowest value of $V_0$ for given  $\Delta_{\rm z,1}$ corresponds to an interface electronic density of $2\cdot 10^{13}/{\rm cm}^2$; it is nearly identical for the three sets. The value of $Q=1e^-/{\rm uc}$ for the LAO polar film is specially marked in the plots. At the point where the system undergoes a transition from $n=0$ to $0<n<Q$, the inverse compressibility displays a sharp drop as a function of $Q$. Note that the differential capacitance is only defined for $0<n<Q$, that is, when a gate bias can transfer charges from the surface to the interface.}
\label{MultibandCompCapQ}
\end{figure*}

In this article we addressed explicitly the dependence of the capacitance of the heterostructure on the RSOC. As reiterated above, the  differential capacitance $C_{\rm diff}$ is not the compressibility in this scheme. The differential capacitance is technically found from the second derivative of the free energy with respect to the charge density $n$ at the interface. So $C_{\rm diff}$ is an internal capacitance of the heterostructure, the capacitance between interface and surface. A 3D conducting film  may be deposited on the surface to have a metallic electrode (gold or YBCO, see Appendix~\ref{sec:3Dlayer}). The kinetic contribution to the capacitance suppresses the capacitance below its geometrical value when exchange effects and RSOC can be neglected. However, we find that a strong RSOC can enhance the capacitance with increasing gate bias to values beyond the geometric capacitance. Hence, experiments can verify a RSOC through a capacitance enhancement.

A 6-band model is more realistic to investigate the capacitance of LAO/STO heterostructures (Sec.~\ref{sec:3band}). Most importantly, it allows to derive a Rashba-like SOC through the coupling of an atomic SOC (at the Ti sites) to the $t_{2g}$ bands in the presence of field controlled inversion symmetry breaking at the interface. The 2-band model is interpreted as an effective model for sufficiently small filling when only one $t_{2g}$ orbital is occupied and the splitting of the bands is still linear in momentum $k$, that is, ``Rashba-like''. However, already for density independent RSOC and low filling, the capacitance as a function of bias voltage $V$ behaves differently than in the 2-band model. The capacitance increases perceivably (Fig.~\ref{MultibandnCompCap}(b)), dashed brown line) for the 6-band model whereas it is constant with respect to $V$ for the 2-band model in the absence of a density dependence of the RSOC. 
In order to find capacitances larger than the geometric capacitance $C_{\rm geom}$, one has to include the density dependence of the band structure. 

However, one has to realize that in the regime of stronger gate bias $V$, a variable number of bands are occupied (see Fig.~\ref{MultibandnCompCap}(a)) and the splitting of the SOC band pairs is not anymore linear in $k$ (see Fig.~\ref{MultibandSplitting}, the middle band pair has anyway no linear splitting, even for $k\rightarrow 0$). Then the capacitance as a function of $V$ depends strongly on the chosen parameter set ($t_{2g}$ band splitting, effective masses, SOC, orbital deformation). We could identify bias-induced capacitance enhancements of the order of 5\% for parameter sets compatible with the LAO/STO heterostructure, but a 30\% enhancement as observed in experiments~\cite{Li} appears to be beyond the present modelling. It should be noted that those experiments are in a regime of very low charge carrier densities and, therefore, it is expected that electronic exchange is much more prominent, an effect which would have to be included. 

For a realistic value of the effective distance between interface and top gate, $d/\varepsilon=10\,a_{\rm B}$, the total compressibility of the system stays positive even for large values of the parameter $\Delta_{\rm z,1}$ that controls the density dependence of the hybridization between the $d_{\rm xy}$ orbital and the $d_{\rm xz,yz}$ orbitals. There is no instability which produces a phase separation on account of a Rashba-like SOC. 

We portray the basic capacative properties of a polar heterostructure in Fig.~\ref{MultibandCompCapQ} by relating the compressibility, the relative value of the interface electronic densitiy and the differential capacitance to a variable polar charge density. The value of $Q=1e^-/{\rm uc}$ for the LAO/STO heterostructure is marked. We assigned $\Delta_{\rm z,1}=33\, e^2 a_{\rm B}^2$ to the largest value of the RSOC that is still compatible with experiments (see Sec.~\ref{sec:3band}). 
The potential $V_0$ controls the shift of the
interface $t_{2g}$-bands with respect to the surface states: the lowest value of $V_0$ (for each of the three sets of curves for given $\Delta_{\rm z,1}$ in Fig.~\ref{MultibandCompCapQ}) adjusts the value of the interface electronic density $n$ to $2\times 10^{13}/{\rm cm}^2$, which is in the physical range of electronic densities. This corresponds to $n/Q\simeq 0.03$ in Fig.~\ref{MultibandCompCapQ}(b). It is obvious from Fig.~\ref{MultibandCompCapQ}(a) that for $Q=1e^-/{\rm uc}$ and for the LAO/STO values of $V_0$, the compressibility is well in a positive range. However, it is conceivable that for other heterostructures with a polarity of $Q = 2e^-/{\rm uc}$ and large RSOC, that is sizeable $\Delta_{\rm z,1}$, a phase separation is attainable with this mechanism and a considerable enhancement of the differential capacitance over the geometrical value can be achieved (cf.~Fig.~\ref{MultibandCompCapQ}(c)).

\section*{Acknowledgements}  
This work was supported by the Deutsche Forschungsgemeinschaft through TRR~80.
We are grateful to J. Mannhart for his continuing support of our research. The authors acknowledge helpful discussions with R.~Ashoori, S.~Graser, M.~Grilli, J.~Junquera, A.~P.~Kampf, Lu Li, and V.~Tinkl.

\appendix

\section{Calculation of the Energies}

\subsection{Single Layers}\label{appEnergySingleLayers}

For a single layer the strength of the RSOC $\alpha_{\rm R}$ depends on external fields only, so that $\alpha_{\rm R}$ is not a function of the density $n$ in that layer. Hence we obtain the relation between number of electrons $N$ and chemical potential via the energy dispersion, Eq. (\ref{rsocdispersion}):
\begin{align}
N&=\sum_{{\bf{k}},\lambda=\pm}\Theta\left(\mu-\epsilon^\lambda(k)\right)\\
&=\frac{A}{2\pi}\int_0^\infty \D k\cdot k\sum_{\lambda=\pm}\Theta\left[\mu-\left(\frac{\left(k-\lambda\pi N_0\alpha_{\rm R}\right)^2}{2\pi N_0}-\zeta\right)\right]\notag
\end{align}
With the abbreviation $\zeta\equiv-2\epsilon_{\rm min}^-=\pi N_0\alpha_{\rm R}^2/2$ and assuming $\mu+\zeta\geq 0$ (otherwise the Heaviside functions and hence the integrals are zero), we get
\begin{align}
N=\frac{A}{2\pi}\int_0^\infty \D k\cdot k\sum_{\lambda=\pm}\Theta\bigg[&\sqrt{2\pi N_0\mu+(\pi N_0\alpha_{\rm R})^2}\notag\\
&-\left|k-\lambda\pi N_0\alpha_{\rm R}\right|\bigg]
\end{align}
In our analysis we have to distinguish between $\mu > 0$ and $\mu<0$, where both bands or one band is occupied, respectively:
\\\underline{$\mu<0$:}
\begin{subequations}
\begin{align}
4\pi n &= \int_{\pi N_0\alpha_{\rm R}-\sqrt{2\pi N_0\mu+(\pi N_0\alpha_{\rm R})^2}}^{\pi N_0\alpha_{\rm R}+\sqrt{2\pi N_0\mu+(\pi N_0\alpha_{\rm R})^2}}\D k\cdot 2k\notag\\
&= 4\pi N_0\alpha_{\rm R}\sqrt{2\pi N_0\mu+(\pi N_0\alpha_{\rm R})^2}\\
\Rightarrow \mu&= \frac{n^2}{2\pi N_0^3\alpha_{\rm R}^2}-\frac{\pi}{2}N_0\alpha_{\rm R}^2=\frac{n^2}{(2N_0)^2\zeta}-\zeta
\end{align}
\underline{$\mu>0:$}
\begin{align}
4\pi n &= \int_0^{\sqrt{2\pi N_0\mu+(\pi N_0\alpha_{\rm R})^2}+\pi N_0\alpha_{\rm R}}\D k\cdot 2k\notag\\
&\quad+\int_0^{\sqrt{2\pi N_0\mu+(\pi N_0\alpha_{\rm R})^2}-\pi N_0\alpha_{\rm R}}\D k\cdot 2k\notag\\
&=4\pi N_0\mu+4\left(\pi N_0\alpha_{\rm R}\right)^2\\
\Rightarrow \mu &=\frac{n}{N_0}-\pi N_0\alpha_{\rm R}^2 = \frac{n}{N_0}-2\zeta
\end{align}
\end{subequations}
The same distinction has to be made for the calculation of the free energy.\\
\underline{$\mu<0$:}
\begin{subequations}
\begin{align}
F_{0}^<&= \sum_{{\bf{k}},\lambda=\pm}\epsilon^\lambda(k)\Theta\left(\mu-\epsilon^\lambda(k)\right)\notag\\
&= \frac{A}{2\pi}\int_{\pi N_0\alpha_{\rm R}-\sqrt{2\pi N_0\mu+(\pi N_0\alpha_{\rm R})^2}}^{\pi N_0\alpha_{\rm R}+\sqrt{2\pi N_0\mu+(\pi N_0\alpha_{\rm R})^2}} \D k\left(\frac{k^3}{2\pi N_0}-\alpha_{\rm R}k^2\right)\notag\\
&= \frac{2A}{3}\left(\pi N_0\alpha_{\rm R}\mu-\pi^2 N_0^2\alpha_{\rm R}^3 \right)\sqrt{2\pi N_0\mu+(\pi N_0\alpha_{\rm R})^2)}\\
&= A\left(\frac{n^3}{6\pi N_0^3\alpha_{\rm R}^2}-\frac{\pi}{2}N_0\alpha_{\rm R}^2n\right)
\end{align}
\end{subequations}
Note that the $\lambda=-1$ term is zero, since the square root $\sqrt{2\pi N_0\mu+(\pi N_0\alpha_{\rm R})^2}$ is always smaller than $\left|k+\pi N_0\alpha_{\rm R}\right|$ for negative $\mu$. For $\mu>0$ this term yields the integral with the smaller upper bound:\\
\underline{$\mu>0$:}
\begin{subequations}
\begin{align}
F_{0}^>&= \frac{A}{2\pi}\int_{0}^{\sqrt{2\pi N_0\mu+(\pi N_0\alpha_{\rm R})^2}+\pi N_0\alpha_{\rm R}}\D k\left(\frac{k^3}{2\pi N_0}-\alpha_{\rm R}k^2\right)\notag\\
&\ +\frac{A}{2\pi}\int_{0}^{\sqrt{2\pi N_0\mu+(\pi N_0\alpha_{\rm R})^2}-\pi N_0\alpha_{\rm R}}\D k\left(\frac{k^3}{2\pi N_0}+\alpha_{\rm R}k^2\right)\notag\\
&= A\left(\frac{1}{2}N_0\mu^2-\frac{\pi^2}{3}N_0^3\alpha_{\rm R}^4\right)\\
&= A\left(\frac{n^2}{2N_0}-\pi N_0\alpha_{\rm R}^2n +\frac{\pi^2}{6}N_0^3\alpha_{\rm R}^4 \right)
\end{align}
\end{subequations}

\section{Electrostatic Energy}\label{appElectrostaticEnergy}

We calculate the electrostatic energy of a system similar to a LAO/STO multilayer, cf. Fig.~\ref{Electrostatic}: before electronic reconstruction positive and negative charged planes alternate in $z$-direction and the lowest layer is both electrical neutral and free of electrical fields. If an electron charge density $-en$ is transferred from the top to the bottom layer (Fig.~\ref{Electrostatic}(a)), an additional field proportional to $n$ is superimposed. Note that for the RSOC only the asymmetric fields are relevant, so that the field of the electrons in the bottom layer does not contribute to RSOC there. Even if there is only a field in the upper half-space, the field important for SOC \textit{in} the plane is the field originating from the surface in the complete space.
\\
The electrostatic energy of a system with $l$ alternating (double) layers can be calculated via the energy stored in the fields: the electric field $E$ of a planar charge density $e\sigma$ (where $e$ is the electronic charge) is given by
\begin{align}
E=\frac{2\pi}{\varepsilon}e\sigma
\end{align}
and the energy per area between two plates with charge density $\pm e\sigma$ and a distance $d^l$ between them is
\begin{align}
F_{\rm{es}}/A=\frac{1}{2}\frac{\varepsilon}{4\pi}\left(\frac{4\pi}{\varepsilon}e\sigma\right)^2d^l=\frac{2\pi e^2d^l}{\varepsilon}\sigma^2.
\end{align}
Hence the energy per area of $l$ double layers with charge density $\pm eQ$ in an external field, which is generated by a charge density $\pm en$, is
\begin{align}\label{electrostaticen}
F_{\rm{es}}/A =\frac{2\pi e^2 (ld_1^l)}{\varepsilon_1}\left( Q-n\right)^2+\frac{2\pi e^2 (ld_2^l)}{\varepsilon_2}n^2
\end{align}
Fig.~\ref{Electrostatic}(b) shows an equivalent circuit for $l=3$.

\section{\label{sec:3Dlayer}Three-dimensional layer on Top of RSOC Layer}

\begin{figure}[t]
\centering
\includegraphics[width=0.48\textwidth]{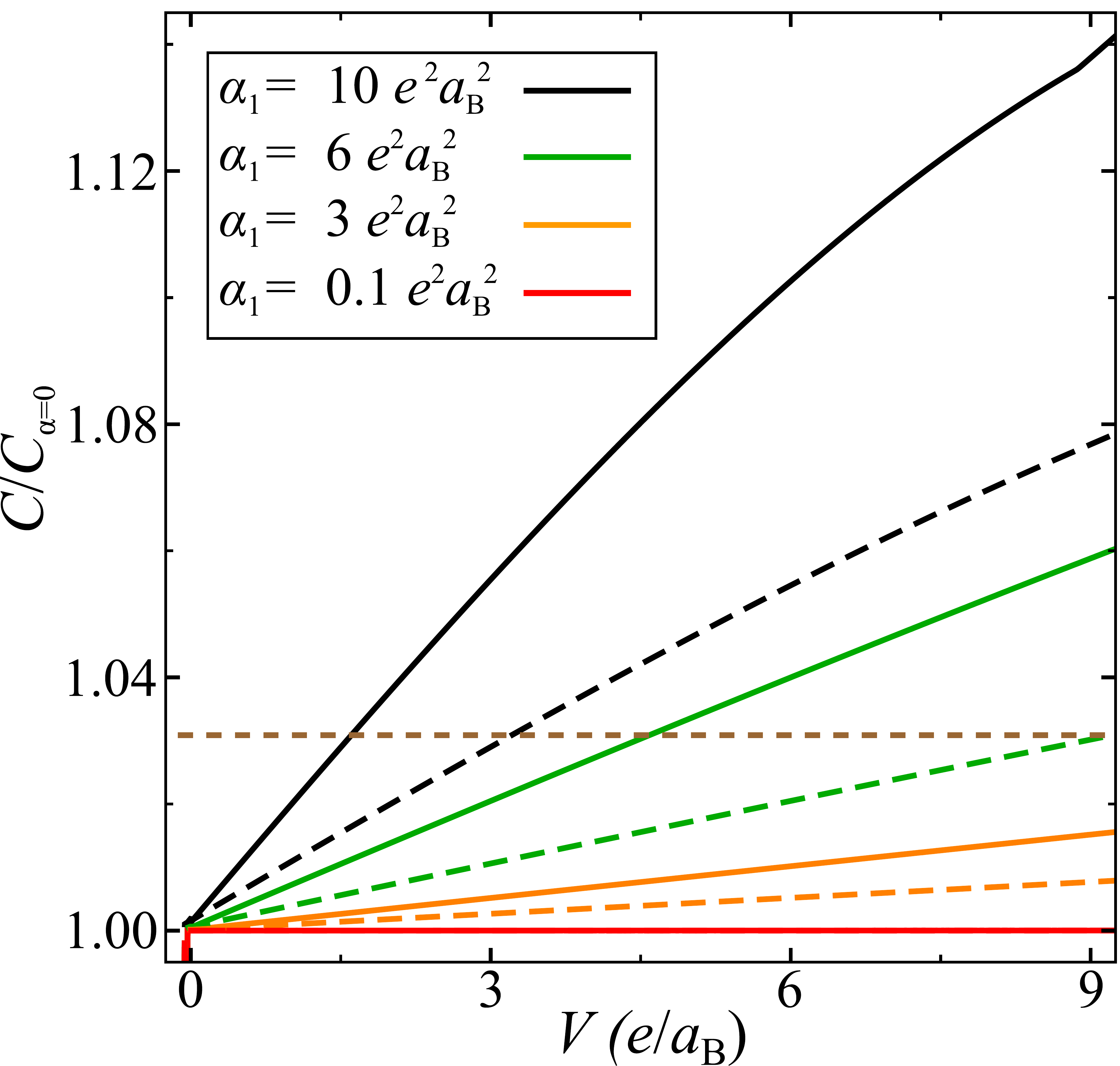}
\caption{(Color online) Differential capacitance per area (solid lines) for a gold electrode  on top of the LAO/STO heterostructure which is terminated by an AlO$_2$ layer. Dashed lines refer to $C=\frac{e\Delta n}{\Delta V}$. Here $m_{\rm STO}=0.7\,m_e$, $d/\varepsilon=10\,a_{\rm B}$, $Q=1e^-/{\rm uc}$, and $\alpha_0=0$. $V_0$ ensures that the density on the interface is $2\cdot 10^{13}\rm{cm}^{-2}$ (see Eq.~(\ref{totalenergy}) for the significance of $V_0$). The gold electrode has a thickness of  $150\,a_{\rm B}$, $m_{\rm Au}=1.1\,m_e$, $\lambda_{\rm TF}=1.57\,a_{\rm B}$ and the capacitance $C_{\rm \alpha=0}/A=6.67\cdot 10^{-3}\,a_{\rm B}^{-1}$ for zero RSOC. The system is in phase P2 for all plotted values of $\alpha_1$, except for the $\alpha_1=10\,e^2a_{\rm B}^2$ curve, where the system is in phase P$1^+$ to the right of the kink at $V=8.8\,e/a_{\rm B}$. The horizontal dashed brown line indicates the geometric capacitance.}
\label{GoldYBCO}
\end{figure}

In some experimental setups, a three-dimensional (3D) top gate is deposited  on top of the LAO/STO heterostructure (see, for example, Ref.~\cite{Li}). The electrode connects with the topmost $\mathrm{AlO_2}$ layer and the electrons therein should be treated as a 3D electron system. 

This setup requires to consider an inhomogeneous system with field penetration in the top electrode as in Ref.~\onlinecite{Buettiker93}. We assume a space independent DOS $N(E_F)$ within the electrode on the ground that this does not modify the qualitative aspects (in fact, modifications are tiny if the electronic system is not close to a van Hove singularity).
Then the kinetic term from the electrode free energy $F_1$ translates into a correction of the effective distance in the geometric capacitance: the effective distance $d/\varepsilon$ is increased by $\lambda_{\rm TF}$, the screening length (see Refs.~\onlinecite{Mead61,Ku64,Simmons65,Stern72,Buettiker93,Black,Kopp}).  
So instead of implementing an inhomogeneous electron system in the energy $F_1$ of the top electrode, one may cast the electrostatic energy into the form (cf.~Eq.~(\ref{mlotorltotalenergy})):
\begin{align} \label{3DF}
F_{\rm{es}}/A = \frac{\pi d e^2}{\varepsilon}n^2 +\pi\left(\frac{d}{\varepsilon}+2\lambda_{\rm TF}\right)e^2\left(Q-n\right)^2.
\end{align}
We assume here that the number of electrons in the 3D film is much larger than the number of unoccupied sites in the interface layer with RSOC, so that even though charge is transferred to the bottom layer, the volume density $n_V$ in the metal remains approximately unchanged. This condition can always be fulfilled by a sufficiently thick metallic layer. Then the screening length is approximately constant---characteristic of the used electrode material. 
          
In Thomas-Fermi theory $\lambda_{\rm TF}$ depends on the volume electron density $n_V$ and the dielectric constant $\varepsilon_c$ of the ion cores in the metal:
\begin{align}
\lambda_{\rm TF}=\sqrt{\frac{\varepsilon_c}{4\pi e^2 N(E_F)}}=\sqrt{\frac{\varepsilon_c E_F}{6\pi e^2 n_V}},
\end{align}
where $N(E_F)$ is the 3D DOS at the Fermi energy. Table \ref{TableofTF} shows the screening lengths for $\mathrm{Au}$, $\mathrm{Ag}$ and $\mathrm{YBa_2Cu_3O_7}$.

\begin{table}[b]
\caption{Material properties and resulting screening lengths $\lambda_{\rm TF}$ for gold, silver and YBCO. Here, 
$n_V$ is the volume electron density and  $\varepsilon_c$ the dielectric constant  of the ion cores in the metal.}
\begin{tabular}{l | c c c c}
 & $E_F$ [${\rm eV}$]& $\varepsilon_c$ & $n_V$ [$10^{21}\,{\rm cm}^{-3}$] & $\lambda_{\rm TF}$\\
\hline
$\mathrm{Au}$ 			& 5.53 \cite{Wyckoff,Ashcroft} 	& 6.9 \cite{Shklyarevskii}	& 59.0 \cite{Wyckoff,Ashcroft} 	& 1.57 		\\
$\mathrm{Ag}$ 			& 5.49 \cite{Wyckoff,Ashcroft}	& 2.5 \cite{Black}	 	& 58.6 \cite{Wyckoff,Ashcroft} 	& 2.99 		\\
$\mathrm{YBa_2Cu_3O_7}$ & 1.0 \cite{Aidam}			& 25  \cite{Aidam} 		& 2\dots 6 \cite{Aidam}		& 7.4\dots 12.8	\\
\end{tabular}\label{TableofTF}
\end{table}

Note that $Q$ and $n_V$ are fixed, as they are determined by the polarity of the LAO layers and the choice of the top gate film. In contrast, the electron density $n$ is found from the minimization of the free energy. 

In Fig.~\ref{GoldYBCO} the differential capacitance $C_{\rm diff}$ and $C=e\Delta n/\Delta V$ are displayed for $\mathrm{Au}$ as top gate. We assumed a screening length of $1.57\,a_{\rm B}$ for gold. For the weakest RSOC, $\alpha_1=0.1\,e^2a_{\rm B}^2$, the capacitance is not altered notably in comparison to the case without RSOC. For the largest value $\alpha_1=10\,e^2a_{\rm B}^2$ (see estimation in Eq.~(\ref{approxalpha1})), we find a capacitance increase of approximately 8~\% in a voltage range  0--100~V. This is the capacitance increase with respect to the case of zero RSOC. The increase with respect to the geometrical capacitance is 5~\% for a gold top electrode.

Different screening lengths cause sizeable differences in magnitude of the capacitances: the screening length in a $\mathrm{YBa_2Cu_3O_7}$  film is $\sim10\,a_{\rm B}$,  and the capacitance increase is approximately 3~\% in the voltage range  0--100~V. The increase with respect to the geometrical capacitance is only 1~\% for the YBCO top electrode.

\section{Relation between compressibility and capacitance and stability criterion}\label{appCompAndCap}

In this section we assume that the electron density at the interface, $n$, does not take one of the boundary values, 0 or $Q$.
Moreover, we assume that the appropriate free energy is minimal with respect to $n(Q,V)$ (see below).

The Helmholtz free energy $F_{\rm{tot}}$ of the system is a function of the total electron density $Q$ and the electron density on the interface $n$ (cf. Eq.~(\ref{totalenergy})):
\begin{align}
F_{\rm{tot}}(Q,n)=F_0(n) +F_1(Q-n)+ F_{\rm{es}}(Q,n) -eV_0nA\notag
\end{align}
The partial derivative of $F_{\rm{tot}}(Q,n)$ with respect to $n$ is the voltage between the plates:
\begin{align}\label{appMinimization}
\frac{\partial F_{\rm{tot}}}{\partial n}=eA\,V(Q,n) 
\end{align}
This thermodynamic relation may be used to identify $n(Q,V)$. We introduce a Legendre transformation of $F_{\rm tot}$ with respect to the variable $n$:
\begin{align}\label{appsecondFV}
F(Q,V) = F_{\rm{tot}}(Q,n(Q,V)) - eA\,V\cdot n(Q,V)
\end{align}
where $F(Q,V)$ depends on the total electron density $Q$ and the external potential $V$ between interface and surface electrode.

The differential capacitance can be derived either from Eq.~(\ref{appMinimization})
\begin{align}\label{appCapacitance}
A/C_{\rm diff}(n)=\frac{\partial V(Q,n)}{e\,\partial n}=\frac{1}{e^2\,A}\frac{\partial^2F_{\rm tot}}{\partial n^2}
\end{align}
where we suppress the label for the variable $Q$ in $C_{\rm diff}$, or from the Legendre transform Eq.~(\ref{appsecondFV}):
\begin{align}\label{appCapacitanceV}
C_{\rm diff}(V)/A=\frac{e\,\partial n(Q,V)}{\partial V}=-\frac{1}{A}\frac{\partial^2 F}{\partial V^2}
\end{align}
Of course, the two capacitances from Eqs.~(\ref{appCapacitance}) and (\ref{appCapacitanceV}) are identical in the sense that $C_{\rm diff}(n(V))=C_{\rm diff}(V)$ holds.

On the other hand, the inverse compressibility $\kappa^{-1}(Q,V)$ in dependence on an external voltage $V$ is the second total derivative of $F$ with respect to the total electron density:
\begin{subequations}
\begin{align}\label{appComp}
\kappa^{-1}\cdot &\,A/Q^2=\frac{\D^2 F(Q,n(Q,V))}{\D Q^2}\\
&=\frac{\D}{\D Q}\left[\partial_Q F+\partial_n F\left(\partial_Q n\right)\right]\notag\\
&=\partial_Q^2 F+2\partial_n\partial_Q F \left(\partial_Q n\right)+\partial_n^2 F\left(\partial_Q n\right)^2\notag\\
&\qquad +\partial_n F\left(\partial_Q^2n\right)\label{appComp2}
\end{align}
\end{subequations}
Note that $\partial_n F \equiv \partial F/\partial n$ is not a derivative with respect to a thermodynamic variable. In fact, the thermodynamic variables are $Q$ and $V$ for $F$. The partial derivative of $F$ with respect to $n$ is identified from the rhs of Eq.~(\ref{appsecondFV}). The thermodynamic potential $F$ is minimal with respect to all values of the internal variable $n$, a necessary condition for thermodynamic stability.

According to Eqs.~(\ref{appsecondFV}) and (\ref{appMinimization}), the last term on the right hand side of Eq.~(\ref{appComp}) is zero. Under a further assumption we can derive more elaborate relations between differential capacitance and compressibility: the dependence of the total free energy $F_{\rm{tot}}$ on the total electron density $Q$ shall be exclusively of the form $Q-n$, so that also
\begin{align}\label{appsecondAssumption}
\frac{\partial^2F}{\partial Q^2} = -\frac{\partial^2 F}{\partial Q\partial n}
\end{align}
is valid. This assumption is reasonable, since $Q-n$ is the electron density in the surface layer, which should be the only density dependent quantity that determines the energy of the surface system. Hence we only demand the electrostatic energy not to have some unusual form. Note that, while $F_{\rm{tot}}(Q,n)=F_{\rm{tot}}(Q-n,n)$ is sufficient for the derivation below, only Eq.~(\ref{appsecondAssumption}) is necessary. Under the validness of this relation the compressibility becomes:
\begin{align}
\kappa^{-1}\,\cdot \,A/Q^2=\frac{\partial^2F}{\partial Q\partial n}\left(2\frac{\partial n}{\partial Q}-1\right)+\frac{\partial^2 F}{\partial n^2}\left(\frac{\partial n}{\partial Q}\right)^2
\end{align}
We use the total derivative of the first condition Eq.~(\ref{appMinimization}),
\begin{align}\label{appDerivofMini}
0&=\frac{\D}{\D Q}\frac{\partial F}{\partial n}\notag\\
&=\frac{\partial^2 F}{\partial Q\partial n} +\frac{\partial^2 F}{\partial n^2}\left(\frac{\partial n}{\partial Q}\right),
\end{align}
to replace $\partial_Q\partial_n F$ in the compressibility above:
\begin{align}\label{appCompdn}
\kappa^{-1}\cdot A/Q^2&=\frac{\partial^2F}{\partial n^2}\left(\frac{\partial n}{\partial Q}\right)\left(1-\frac{\partial n}{\partial Q}\right)\notag\\
&=\frac{e^2A^2}{C_{\rm diff}}\frac{\partial n}{\partial Q}\left(1-\frac{\partial n}{\partial Q}\right)
\end{align}
This relation between the compressibility and differential capacitance can be further analyzed to deduce a stability criterion, i.e., the condition for positive compressibility. Since we assumed that the density $n$ obtained from Eq.~(\ref{appMinimization}) yields a minimum of the free energy,
\begin{align}
\frac{\partial^2 F}{\partial n^2} > 0,
\end{align}
the differential capacitance $C_{\rm diff}$ is positive, which is a necessary criterion for the stability. Note that the second partial derivatives of $F$ and $F_{\rm tot}$ (with respect to $n$ and $Q$) are equal. Hence there is an addtional condition to keep the compressibility positive:
\begin{align}
\kappa>0 \quad\Longleftrightarrow\quad 0 < \frac{\partial n}{\partial Q} <  1.
\end{align}
The system is unstable if adding of electrons causes either electrons to flow from the interface to the surface ($\partial_Q n<0$) or more than the added electrons to flow to the interface ($\partial_Q n>1$). As we will show below, the first alternative does not occur in our model. $\partial_Q n<0$ is in principle the reverse case of $\partial_Q n>1$, i.e., when the isolated surface system would display a negative compressibility.\\
Furthermore, Eq.~(\ref{appDerivofMini}) can be solved for $\partial_Q n$,
\begin{align}
\frac{\partial n}{\partial Q}=-\frac{\partial^2 F}{\partial Q\partial n}\bigg/\frac{\partial^2F}{\partial n^2}.
\end{align}
This result inserted into the compressibility Eq.~(\ref{appCompdn}) yields:
\begin{subequations}
\begin{align}
\kappa^{-1}\cdot A/Q^2&=-\frac{\partial^2 F}{\partial Q\partial n}\left(1+\frac{\partial^2 F}{\partial Q\partial n}\bigg/\frac{\partial^2F}{\partial n^2}\right)\\
&=\frac{\partial^2F}{\partial Q^2}\left(1-\frac{\partial^2F}{\partial Q^2}\bigg/\frac{\partial^2F}{\partial n^2}\right)
\label{appCompdiff}\\
&=\frac{\partial^2F}{\partial Q^2}\left(1-\frac{C_{\rm diff}}{e^2A^2}\frac{\partial^2F}{\partial Q^2}\right)
\end{align}
\end{subequations}
We consider a system where the electrons in the surface electrode are not subject to a RSOC or other effects which may produce a negative compressibility. Correspondingly, we may assume that 
$
{\partial^2 F_1}/{\partial Q^2} > 0
$.
As the electrostatic contribution to the free energy also generates a positive second derivative with respect to $Q$, the free energy $F$ has 
to obey
\begin{align}\label{appF2Q}
\frac{\partial^2 F}{\partial Q^2} > 0.
\end{align}
Since $\partial_n\partial_Q F$ is negative (Eq.~(\ref{appsecondAssumption}) with relation (\ref{appF2Q})) and $\partial^2F/\partial n^2$ is positive in the considered regime, Eq.~(\ref{appDerivofMini}) shows that $\partial_Q n$ has to be positive in our model. 

We consider the regime of positive capacitance---is it then possible to identify a negative compressibility of the complete electronic system?  In fact, according to Eq.~(\ref{appCompdiff}) and the positivity of $\partial^2F/\partial Q^2$, the compressibility is negative if
\begin{align}\label{appNegComp}
\kappa<0\,&\Longleftrightarrow\; \frac{\partial^2F}{\partial n^2}<\frac{\partial^2F}{\partial Q^2}\notag\\
 &\Longleftrightarrow\; \frac{\partial^2F_{\rm tot}}{\partial n^2}<\frac{\partial^2F_{\rm tot}}{\partial Q^2}\notag\\
&\Longleftrightarrow\; \frac{\partial^2(F_0+F_1)/A}{\partial n^2} + 4\frac{\pi de^2}{\varepsilon}
   < \frac{\partial^2F_1/A}{\partial n^2} + 2\frac{\pi de^2}{\varepsilon}\notag\\
&\Longleftrightarrow\; \frac{\partial^2 F_0/A}{\partial n^2} + 2\,\frac{\pi de^2}{\varepsilon} < 0 
\end{align}
This inequality does not involve properties of the surface system explicitly. With the assumption that $C_{\rm diff}$ is positive, one finds that the compressibility of the heterostructure is negative, if the sum of the inverse compressibility of the interface system (times $n^2$) and half the inverse geometric capacitance $2\pi de^2/\varepsilon$ is smaller than zero. 

We emphasize that the inverse capacitance is a sum, the terms of which are generated by the second derivatives of different contributions to the Helmholtz free energy $F_{\rm{tot}}$ in Eq.~(\ref{appCapacitance}), consistent with previous analyses (e.g., Ref.~\cite{Kopp}). The inverse compressibility is not the sum of these term: it is the asymmetric setup that causes $(C_{\rm diff}/A)^{-1} \neq  \kappa^{-1}/ (e^2 Q^2)$. However, the relation (\ref{appNegComp}) contradicts the identification of a negative compressibility in Refs.~\cite{Grilli,Bucheli}: there the authors missed the ``electrostatic compressibility'' $2\pi de^2/\varepsilon$ entirely and the compressibilty of the interface electronic system was evaluated differently (see~Appendix~\ref{appEnergyVsChemPot}).

In our model the total free energy $F_{\rm tot}$ is of the form presented in Eq.~(\ref{totalenergy}), which reads with the abbreviation $D\equiv\pi de^2/\varepsilon=(e^2/4)\cdot A/C_{\rm geom}$:
\begin{align}
F_{\rm tot}(Q-n,n)&= F_0(n)+\left(\frac{1}{2N_1}+D\right)(Q-n)^2A\notag\\
&\qquad+Dn^2A-eV_0 n A.\label{appFtot}
\end{align}
Due to the analytical form of $\partial_Q^2F_{\rm tot}=(1/N_1+2D)A$, Eq.~(\ref{appCompdiff}) yields a direct relation between the measured differential capacitance and the total compressibility:
\begin{align}\label{appKappaC}
\frac{1}{Q^2}\,\kappa^{-1}=\left(\frac{1}{N_1}+2D\right)\left[1- \frac{C_{\rm diff}}{e^2A}\left(\frac{1}{N_1}+2D\right)\right]
\end{align}
and the transfer of charge to the interface is:
\begin{align}\label{appdndQ}
\frac{\partial n}{\partial Q}=\left(\frac{1}{e^2 N_1}+\frac{1}{2}\frac{A}{C_{\rm geom}}\right)\,\frac{C_{\rm diff}}{A}\,.
\end{align}
Obviously, for the symmetric case without RSOC and with equal DOS at the interface and surface electrode, $N_0=N_1$, one finds ${\partial n}/{\partial Q}=\frac{1}{2}$ and, consequently, $n^2 \kappa = C_{\rm diff}/(e^2 A)$.

For effective interface-surface distances $d/\varepsilon \gtrsim 10\, a_{\rm B}$, the electrostatic term $2D=\frac{e^2}{2}\cdot A/C_{\rm geom}$ dominates over the inverse
compressiblities $1/N_{0,1}$, and the charge transfer ${\partial n}/{\partial Q}$ is close to $\frac{1}{2}$ for vanishing RSOC. Therefore, sizeable deviations of ${\partial n}/{\partial Q}$ from $\frac{1}{2}$ can be indicative of a substantial RSOC in these systems.
With the measurement of $C_{\rm diff}$ and knowledge of $C_{\rm geom}$ and $N_1$ one can determine ${\partial n}/{\partial Q}$ and $\kappa$ through Eqs.~(\ref{appdndQ}) and (\ref{appKappaC}), respectively.

\begin{figure}[t]
\centering
\includegraphics[width=0.9\columnwidth]{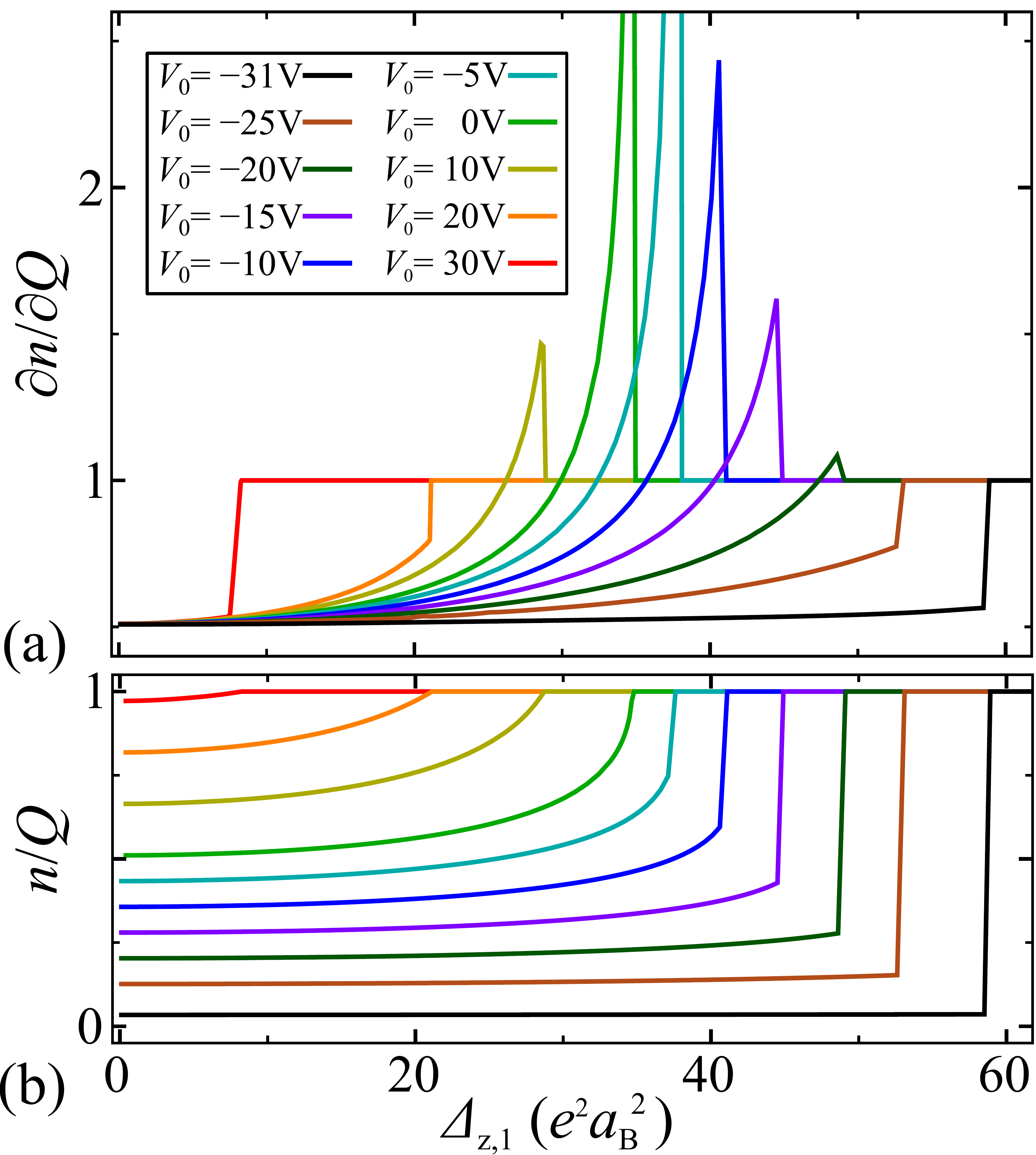}
\caption{(Color online) Differential charge transfer $\partial n/\partial Q$ (a), and the corresponding interface electronic density $n$ (b), both with respect to $\Delta_{\rm z,1}$, which tunes the density dependence of the RSOC. The density $n$ is controllel by $V_0$. Here, $m_{\ell},=0.7\,m_{\rm e}$, $m_{\rm h}=15\,m_{\rm e}$, $m_1=m_{\rm e}$, $d/\varepsilon=10\,a_{\rm B}$, $Q=1\,e^-/{\rm uc}=6.6\cdot 10^{14}$/cm$^2$, $\Delta_{\rm SO}=10\,{\rm meV}$, $\Delta_{\rm E}=50\,{\rm meV}$, and $\Delta_{\rm z,0}=10^{-3}\,e^2$. We find $C_{\rm diff}>C_{\rm geom}$ for $\partial_Q n > 41/80$. The jump to the horizontal line with $\partial n/\partial Q=1$ in (a) takes place when all electronic charge is accumulated at the interface, see (b).}
\label{MultibanddndQ}
\end{figure}

The electronic transfer ${\partial n}/{\partial Q}$ is displayed in Fig.~\ref{MultibanddndQ} as a function of the density dependent hybrization, which effectively controls the strength of the RSOC (see Sec.~\ref{sec:3band}). The limiting value of ${\partial n}/{\partial Q}=1$ is reached, when for large RSOC all additional charge carriers are accumulated at the interface (horizontal lines in Fig.~\ref{MultibanddndQ}).

\section{Compressibility for the phase with charge-depleted surface ($Q=n$)}\label{appEnergyVsChemPot}

The first derivative of the energy with respect to the electron density $n$ yields the chemical potential while the second derivative generates the inverse compressibility $\kappa^{-1}$. We take the strength of the RSOC $\alpha_{\rm R}$ density dependent, and the density dependence is implemented already in the free energy.  However, Caprara {\it et al.}~\cite{Grilli} determine the chemical potential from a model with given $\alpha_{\rm R}$, then introduce the density dependence of $\alpha_{\rm R}$ at that level and calculate the inverse compressibility $\kappa_{\rm C}^{-1}$ from the first derivate of the chemical potential with respect to $n$.

Here we compare the resulting compressibilities $\kappa$ and $\kappa_{\rm C}$ of these two approaches. Therefor we use the abbreviation $\zeta\equiv\pi N_0\alpha_{\rm R}^2/2$, which equals the lowest energy in the dispersion Eq. (\ref{rsocdispersion}), and rewrite the energy Eq. (\ref{mlotorlrashbaenergy}) of a layer with RSOC:
\begin{align}\label{appEVCPEnergy}
F_0^\lessgtr(n)/A= \left\{ \begin{array}{cl}\displaystyle \frac{n^3}{12 N_0^2\zeta}-\zeta n &,\mu<0\\
\displaystyle \frac{n^2}{2N_0}-2\zeta n +\frac{2}{3}N_0\zeta^2 &,\mu>0 \end{array}\right.
\end{align}
For density independent $\alpha_{\rm R}$ and hence density independent $\zeta$, the chemical potential $\mu$ is given by
\begin{align}
\mu_{\rm C}^\lessgtr=\frac{\D F_0^\lessgtr}{A\D n}= \left\{ \begin{array}{cl}\displaystyle \frac{n^2}{4 N_0^2\zeta}-\zeta &,\mu<0\\
\displaystyle \frac{n}{N_0}-2\zeta &,\mu>0 \end{array}\right. .
\end{align}
Caprara {\it et al.}~\cite{Grilli} then set $\alpha_{\rm R}=\alpha_{\rm R}(n)$, and their inverse compressibility is
\begin{align}
\left(\kappa_{\rm C}^\lessgtr\right)^{-1}=n^2\frac{\D \mu_{\rm C}^\lessgtr}{\D n}=n^2\left\{ \begin{array}{cl}\displaystyle \frac{n}{2 N_0^2\zeta}-\zeta'\left[\left(\frac{n}{2N_0\zeta}\right)^2+1\right]\\
\displaystyle \frac{1}{N_0}-2\zeta' \end{array}\right.
\end{align}
Here $\zeta'= d\zeta/dn$ and $\zeta''= d^2\zeta/dn^2$.
In contrast, if we set $\alpha_{\rm R}=\alpha_{\rm R}(n)$ already in the energy $F_0^\lessgtr$ in Eq. (\ref{appEVCPEnergy}):
\begin{align}
\mu^\lessgtr=\frac{\D F_0^\lessgtr}{A\D n}= \left\{ \begin{array}{cl}\displaystyle \mu_{\rm C}^<-\zeta'\left[\frac{n^3}{12N_0^2\zeta^2}+n\right] &,\mu<0\\
\displaystyle \mu_{\rm C}^>-\zeta'\left[2-\frac{4}{3}N_0\zeta\right] &,\mu>0 \end{array}\right. .
\end{align}
\begin{subequations}
\begin{align}
(\kappa^<)^{-1}=(\kappa^<_{\rm C})^{-1}&+n^2\left(\frac{n^2}{2N_0^2\zeta^2}\left(\frac{n(\zeta')^2}{3\zeta}-\frac{\zeta'}{2}-\frac{n\zeta''}{6} \right) \right)\notag\\
&-n^2\left(\zeta'-\zeta''n\right)
\end{align}
\begin{align}\label{kappaC}
(\kappa^>)^{-1}=(\kappa^>_{\rm C})^{-1}&+n^2\left(-2\zeta'+\frac{4}{3}N_0\left((\zeta')^2+\zeta\zeta''\right)-2n\zeta'' \right)
\end{align}
\end{subequations}
For $\alpha_{\rm R}=\alpha_1 n$ this simplifies to
\begin{align}
\left(\kappa_{\rm C}^\lessgtr\right)^{-1}=n^2\left\{ \begin{array}{cl}\displaystyle -\pi N_0 \alpha_1^2n  &,\mu<0\\
\displaystyle \frac{1}{N_0}-2\pi N_0\alpha_1^2n &,\mu>0\end{array}\right.
\end{align}
and
\begin{align}\label{kappaS}
\left(\kappa^\lessgtr\right)^{-1}=n^2\left\{ \begin{array}{cl}\displaystyle -3\pi N_0 \alpha_1^2n &,\mu<0\\
\displaystyle \frac{1}{N_0}-6\pi N_0\alpha_1^2n +2\pi^2N_0^3\alpha_1^4n^2 &,\mu>0\end{array}\right.
\end{align}
As one can see from Eqs.~(\ref{kappaC})  and (\ref{kappaS}), $(\kappa^<_{\rm C})^{-1}$ and $(\kappa^<)^{-1}$ differ by a factor of 3. The deviation for $(\kappa^>)^{-1}$ is less since $1/N_0$ dominates (\ref{kappaC})  and (\ref{kappaS}) for realistic values of $n$, $\alpha_1$, and $N_0$.


\begin{references}
\bibitem{comment1} The charge density, which generates the electric field in the RSOC, is not the interface charge carrier density $n$ --- the latter produces a symmetric electric field which does not entail a RSOC. It is rather the compensating charge in the LaAlO$_3$ layers and at the LaAlO$_3$/SrTiO$_3$ surface that generates the field which controls the RSOC in the interface layer. One might be lead to the conclusion that this is a minor detail as the compensating or ``external'' charge is of the same magnitude as $n$. However, the thermodynamic expression for the compressibility does not allow to vary the external charge and, therefore, a negative compressibility seems to be difficult to realize in this scheme.
\end{references}
\end{document}